\documentclass[msom]{informs3}

\usepackage{makecell}
\usepackage{siunitx}
\usepackage{tablefootnote}
\usepackage[flushleft]{threeparttable}
\usepackage{tikz}  
\usepackage{xurl}
\usepackage[colorlinks = true,
            linkcolor = red,
            urlcolor  = blue,
            citecolor = blue]{hyperref}
\usetikzlibrary{shapes.multipart}
\OneAndAHalfSpacedXI 
 
\usepackage{enumitem}
\usepackage{xcolor}
\definecolor{darkblue}{HTML}{0F07AD}
\usepackage{physics}
\usepackage{endnotes}
\usepackage{mathtools}
\usepackage{bbm}
\usepackage{pdflscape}
\usepackage{subfig} 
\usepackage{booktabs}
\usepackage[toc,page]{appendix}
\usepackage[]{algorithm2e}
\SetKwRepeat{Do}{do}{while}
\SetKw{KwBy}{by}
\SetKw{KwFrom}{from}
\SetKw{KwContinue}{continue}
\SetKw{KwBreak}{break}
\renewcommand{\qed}{$\square$}

\usepackage[colorlinks=true,citecolor = blue]{hyperref} \let\footnote=\endnote
 \usepackage[font=scriptsize,labelfont=bf]{caption}
\usepackage{bbm}
\let\footnote=\endnote

\usepackage{tikz}   
\usetikzlibrary{chains,shapes.multipart} 
\usetikzlibrary{shapes,calc,fit}
\usetikzlibrary{automata,positioning}

\tikzstyle{status} = [rectangle, draw=black, text centered, anchor=north,
text=black, minimum width=2em, minimum height=2em, node  distance=6ex and 7em,
font=\bfseries] \tikzstyle{line} = [draw,thick,-latex] \tikzstyle{transition} = [font=\small]
\tikzstyle{transition} = [font=\small] 
\tikzset{
paths1/.style={->,dashed}, 
paths/.style={->},
paths2/.style={<->},
path/.style={-},
queuei/.pic={
  \draw[line width=1pt]
    (-0.3,0) -- ++(2cm,0) -- ++(0,-1cm) -- ++(-2cm,0);
   \foreach \Val in {1,...,3}
     \draw ([xshift=-\Val*10pt]1.5cm,0) -- ++(0,-1cm);
   \node[above] at (1cm,0) [] {};
  },
}  
  
\usepackage{natbib}
\bibpunct[, ]{(}{)}{;}{a}{,}{,}

 \def\bibfont{\footnotesize}%
\TheoremsNumberedThrough     
\ECRepeatTheorems

\EquationsNumberedThrough    

\begin{document}


\RUNAUTHOR{Eugene Furman, Arik Senderovich, Shane Bergsma, J. Christopher Beck}

\RUNTITLE{Capacity Allocation for Clouds with Parallel
Processing, Batch Arrivals, and Heterogeneous
Service Requirements}

\TITLE{Capacity Allocation for Clouds with Parallel Processing, Batch Arrivals, and Heterogeneous Service Requirements}
 
\ARTICLEAUTHORS{%
\AUTHOR{Eugene Furman}
\AFF{Operations and Decision Sciences, Alba Graduate Business School, American College of Greece,
Athens, Greece\\
\EMAIL{efurman@alba.acg.edu}}
\AUTHOR{Arik Senderovich}
\AFF{School of Information Technology,
York University, Toronto,
Ontario\\
\EMAIL{sariks@yorku.ca}} 
\AUTHOR{Shane Bergsma}
\AFF{Toronto Research Center,
Huawei Cloud, Markham,
Ontario\\
\EMAIL{shane.bergsma@huawei.com}}
\AUTHOR{J. Christopher Beck}
\AFF{Department of Mechanical \& Industrial Engineering,
University of Toronto, Toronto,
Ontario\\
\EMAIL{jcb@mie.utoronto.ca}}
} 

\ABSTRACT{\textbf{Problem Definition:} Allocating sufficient capacity to cloud services is a challenging task, especially when demand is time-varying, heterogeneous, contains batches, and requires multiple types of resources for processing. In this setting, providers decide whether to reserve portions of their capacity to individual job classes
or to offer it in a flexible manner.
\textbf{Methodology/results:} In collaboration with Huawei Cloud, a worldwide provider of cloud services, we propose a heuristic policy that allocates multiple types of resources to jobs and also satisfies their pre-specified service level agreements (SLAs). We model the system as a multi-class queueing network with parallel processing and multiple types of resources, 
where arrivals (i.e., virtual machines and containers) follow time-varying patterns and require at least one unit of each resource for processing. While virtual machines leave if they are not served immediately, containers can join a queue. We introduce a diffusion approximation of the offered load of such system and investigate its fidelity as compared to the observed data. Then, we develop a heuristic approach that leverages this approximation to determine capacity levels that satisfy probabilistic SLAs in the system with fully flexible servers. \textbf{Managerial Implications:} Using a data set of cloud computing requests over a representative 8-day period from Huawei Cloud, we show that our heuristic policy results in a 20\% capacity reduction and better service quality as compared to a benchmark that reserves resources. In addition, we show that the system utilization induced by our policy is superior to the benchmark, i.e., it implies less idling of resources in most instances. 
Thus, our approach enables cloud operators to both reduce costs and achieve better performance. 
} 


\maketitle
\section{Introduction}\label{Sec:intro}
Remote operational processes
require large quantities of computing resources such as Central Processing Units (CPUs), Random Access Memory (RAM), etc. By 2025, the amount of data generated worldwide is to exceed 200 zettabytes with more than a half of it to be processed by cloud platforms~\citep{EnterpriseIT,gartner,Cloudwards}. 

Often, Virtual Machines (VMs), i.e., software-based remote computers within another computer's operating system, deliver remote applications and services to end users \citep{Cloudflare,Cloudwards}. Once hardware resources are provisioned to a VM, its users are permitted to run applications in real time as if they were stored on their local systems. Unfortunately, VMs 
require expertise to set up and manage. Alternatively, 
containerized applications, i.e., lightweight software packages with all the dependencies required for their execution, have been gaining popularity as they may run directly on the server~\citep{Atlassian}. Notice that although we model containers as running directly on the server, cloud providers may execute containers inside special lightweight ``microVMs'' in order to ensure security isolation \citep{verma2015large,cortez2017resource}.

In 2020, a Cloud Native Computing Foundation survey found that the use of containers had grown 300\% since 2016~\citep{CNCF}. Later, in 2021, the International Data Corporation highlighted an ongoing operational shift of the cloud computing industry to a paradigm where 80\% of global cloud computing needs will be met by containerized solutions~\citep{CIO}. According to Gartner, the usage of containers across enterprises is already on the rise as 90\% of global organizations are to run containerized applications by 2026 (up from 40\% in 2021)~\citep{Gartner2021}. 

Unsurprisingly, to satisfy strict service-level agreements (SLAs), VMs and containers require cloud computing infrastructure with near zero-latency capabilities that has become an essential commodity for many corporations \citep{bruckner2002capturing}. In particular, our partner organization, Huawei Cloud, a provider of a wide range of cloud-computing services worldwide, would like to complement
their VM-based platforms with containerized solutions, such that 
the underlying hardware resources can be dynamically shared without a necessity to dedicate them to VMs or containers. 
However, effective deployment of the fully flexible resources requires smart capacity allocation algorithms that allow providers to adapt to the dynamically changing demand.  

Motivated by the challenge of setting capacity levels in a time-varying environment, we consider operations of Huawei Cloud which offers remote computing functionality for internal company use and to external subscribers on-demand. Any dynamic provisioning of capacity is linked to optimal determination of quantities of CPU and RAM units required to satisfy heterogeneous SLAs of VMs and multiple classes of containers to be running jointly on a cloud platform. 

We represent an aggregate capacity of the company's cloud computing infrastructure by the total number of homogeneous CPU cores, i.e., individual processors within a CPU, and gigabytes of memory. 
Without loss of generality, we assume that CPU cores are the most constraining resource with a higher cost of shortages (i.e., dominant). Further, VMs and containers arrive stochastically to the system; they simultaneously require multiple CPU cores and units of memory with a random service duration associated with their size. Typically, VMs are large infinitely impatient jobs, (i.e., they must begin processing upon arrival to the system)
and may spend up to several years in service. In contrast, containers join a queue if needed, and their service times rarely exceed 8 hours.

To capture the non-stationarity in demand, we assume jobs arrive according to a compound non-homogeneous Poisson process, i.e., batch arrivals are allowed. Resource requirements and service durations are generally distributed. Our objective is to determine the minimum non-stationary capacity (i.e., quantity of CPU cores and units of RAM) such that pre-specified service level agreements are satisfied over a planning horizon. To the best of our knowledge, exact analysis of the time-varying stochastic system with batches and multiple resources is intractable, and thus, we develop a three-stage procedure that sets the desired capacity levels heuristically. In the first stage (the $\infty$ stage), we determine the probability distribution of the system's stochastic offered load (SOL) and propose a corresponding diffusion model. In the second stage (the $\alpha$ stage), by collapsing all jobs into a single fictitious class and by adapting a square-root-staffing rule, we set a capacity level for the dominant resource. Finally, in the last stage (the $\epsilon$ stage), we employ the square-root-staffing rule again to determine quantities of non-dominant resources.  

Using a data set of cloud computing requests for VMs over a representative period of 8 days, 
provided to us by Huawei Cloud, we motivate our modelling choices by showing that demand for VMs is time-varying and exhibits periodicity throughout the course of a single week. We also confirm that batch arrivals of requests for VMs can be modelled by a non-homogeneous Poisson process reasonably well. Further, we conduct a simulation study to demonstrate the practicality of our methodology. We first verify that the 80-percentile interval of the generated stochastic paths of the SOL confines the CPU-core workload in the infinite capacity system parameterised by the observed data. 
Then, using a data-driven parameterization, we generate workload for a system with 5 classes of jobs (i.e., VMs and four types of containers) with heterogeneous service requirements. We apply our three-stage procedure to set capacity levels in such system.
In particular, we perform a set of near-stationary and time-varying experiments that compare performance of our heuristic policy with a benchmark, 
which emulates Huawei Cloud's existing operations by dedicating capacity to job classes~ \citep{ke2021fundy}. We find that, in both regimes, the proposed heuristic outperforms the benchmark by requiring at least 20\% fewer resources while respecting pre-specified SLAs. Thus, Huawei Cloud can potentially reduce their aggregate cloud computing capacity while also improving quality of their service by sharing resources instead of dedicating them to various types of workload.

We contribute to the operations research literature by extending previous work on non-stationary staffing policies for queueing systems with parallel processing, batch arrivals, and time-varying, periodic demand. Our approach of combining a diffusion model with a state-of-the-art square-root-staffing rule can be applied to many systems, where practitioners are able to determine which resource is dominant; it can also accommodate a broad class of arrival functions that are general enough to describe most periodic demand processes in service settings. In addition, our methodology is computationally efficient as it leverages closed-form expressions for the SOL. Thus, our service policy can dynamically adapt to changes in demand and remains deployable in practice.  

\section{Motivation: Exploratory Data Analysis}\label{sec:EDA}
Huawei Cloud delivers services to consumers around the world. Although each service has a specific business focus,
the typical demand is heterogeneous, time-varying, and contains batches (i.e., jobs can be submitted for processing at the same time). 
Our data set includes arrival times, service durations, and capacity requirements of requests for virtual machines (VMs) over a period of 8 days 
for one such service. It accounts for 196,770 arrivals where the service times range from approximately 10 milliseconds to 8 days. Each VM requires at least 1 CPU core and 1 gigabyte of memory. In this section, we explore this data set to motivate our modelling choices. Specifically, we visualize the service requirements of jobs and demonstrate that jobs belonging to the same batch have low variability in their service times. We also show that the observed arrivals are time-varying and periodic with a period of 1 day. Further, we confirm that counts of job batches arriving over a weekday satisfy the statistical non-homogeneous Poisson test as described in \cite{brown2005statistical}. 
\subsection{Descriptive Summary of Service Requirements}\label{sec:dscSum}
\begin{figure}[h] 
\caption{Service Requirements: (a) CPU cores; and (b) RAM in Gigabytes.}
\label{resHist}
     \subfloat[]{%
        \includegraphics[width=0.53\textwidth]{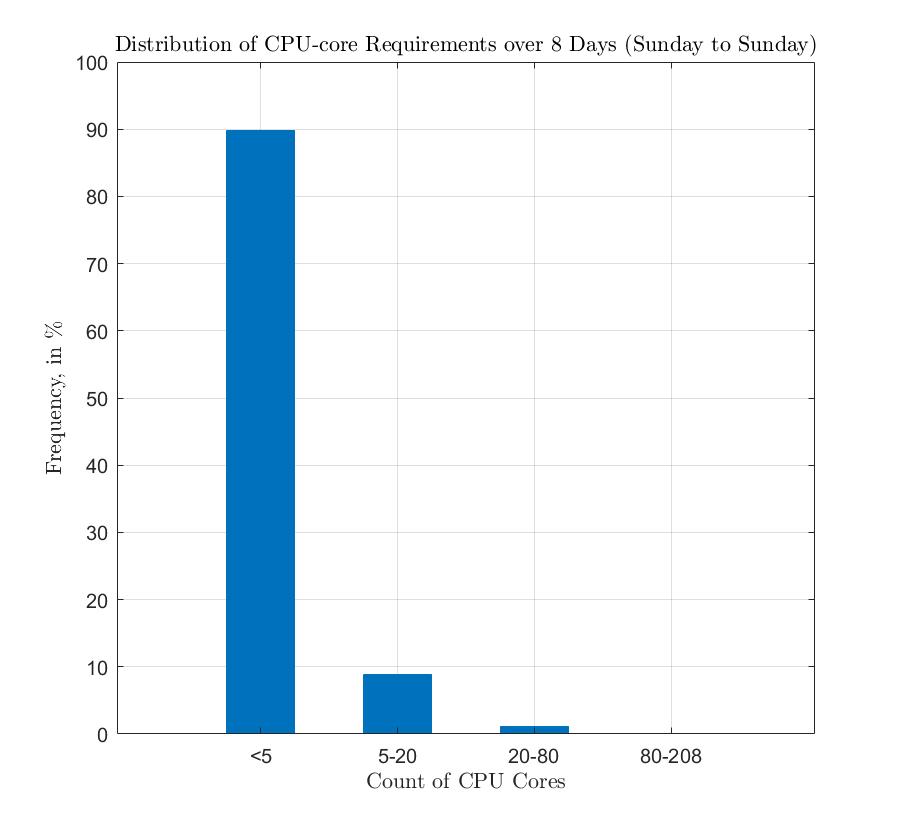}%
        \label{cpuReq}%
         }%
     \hfill%
     \subfloat[]{%
        \includegraphics[width=0.53\textwidth]{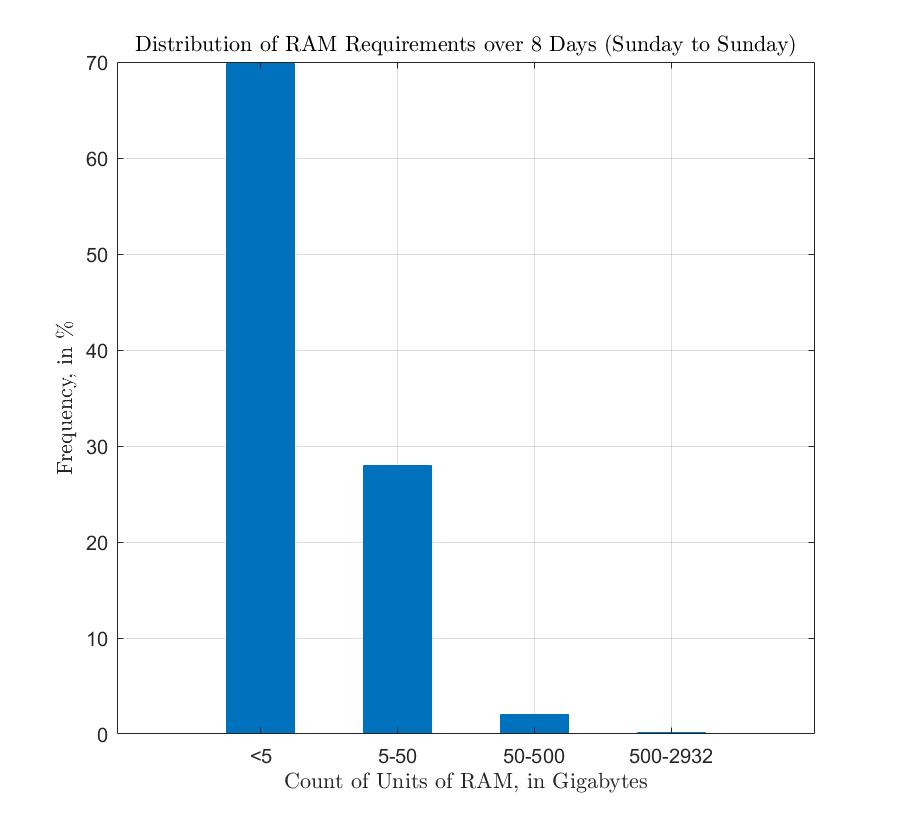}%
        \label{ramReq}
         }
\end{figure}

To summarize resource requirements of jobs in our data, we present their distributions in Figure~\ref{resHist}, where the x-axes describe counts of resources, while the y-axes correspond to their frequencies in percentage.    
According to Figure~\ref{cpuReq}--\ref{ramReq}, 90\% of VMs require less than 5 CPU cores for processing, while approximately 70\% of VMs require less than 5 gigabytes of memory. However, less than 1\% of jobs need as many as 200 CPU cores, and/or 3000 gigabytes of RAM for their service.   

\begin{figure}[h] 
\caption{Summary: (a) Service Times; (b) Variability of Service Times amongst Jobs in the Same Batch.}
\label{sTimesHist}
     \subfloat[]{%
        \includegraphics[width=0.53\textwidth]{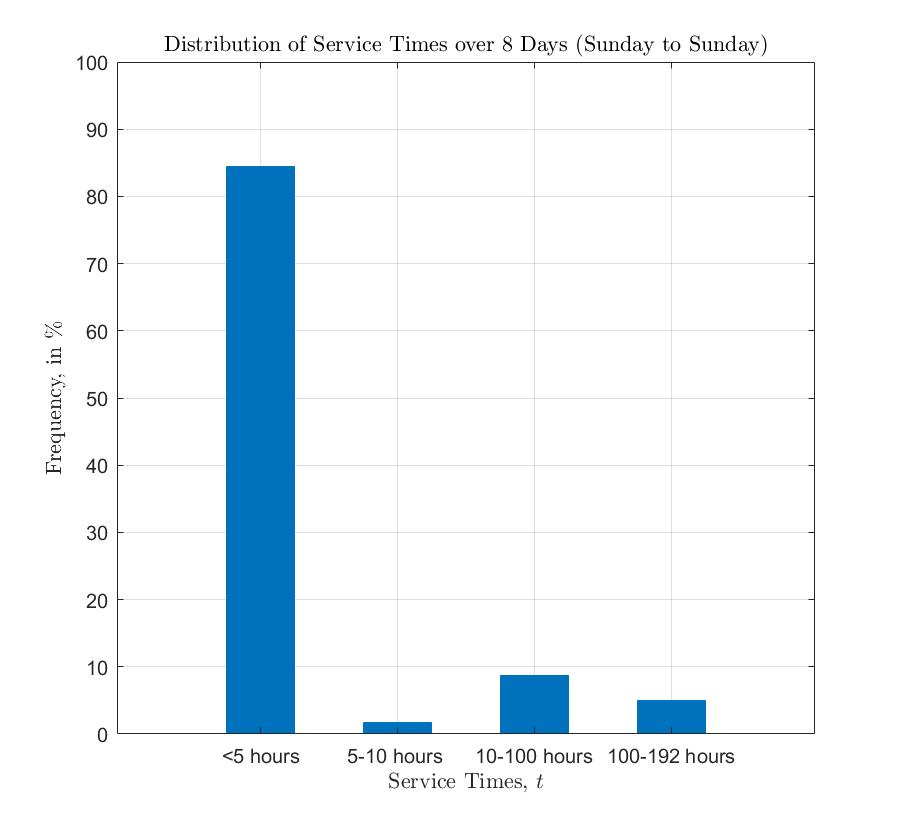}%
        \label{sTimesH}%
         }%
     \hfill%
     \subfloat[]{%
        \includegraphics[width=0.53\textwidth]{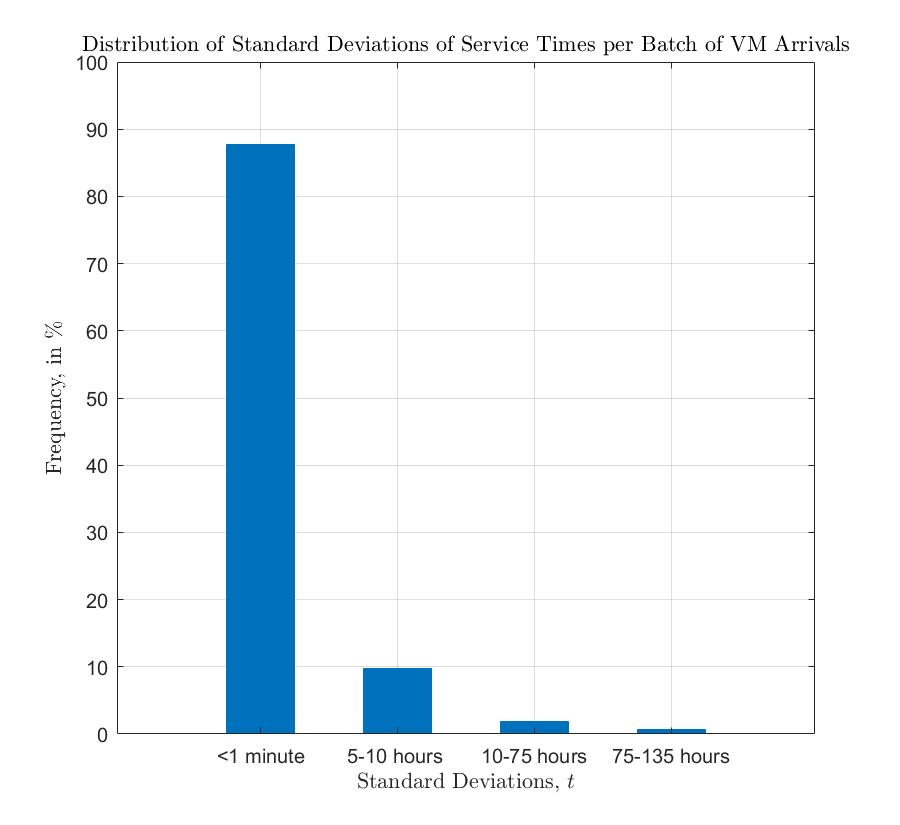}%
        \label{sTimesVar}
         }
\end{figure}

Our data set is characterized by a large fraction of jobs with short service durations. According to Figure~\ref{sTimesH}, almost 85\% percent of VMs require less than 5 hours to get serviced, and only a few jobs occupy cloud computing capacity for as long as 7 days. Further, our data set accounts for the total of 123,815 batches of jobs. We summarize variability in service times of jobs belonging to the same batch in Figure~\ref{sTimesVar}, which represents a frequency plot of their standard deviations. According to this figure, these standard deviations do not exceed 1 minute in nearly 90\% of instances.

\subsection{Understanding the Arrival Process}\label{sec:arrPro}
\begin{figure}[h] 
\caption{Arrival Process: (a) Histogram of Arrival Times; (b) Non-homogeneous Poisson Process Test.}
\label{arrTimesPlots}
     \subfloat[]{%
        \includegraphics[width=0.53\textwidth]{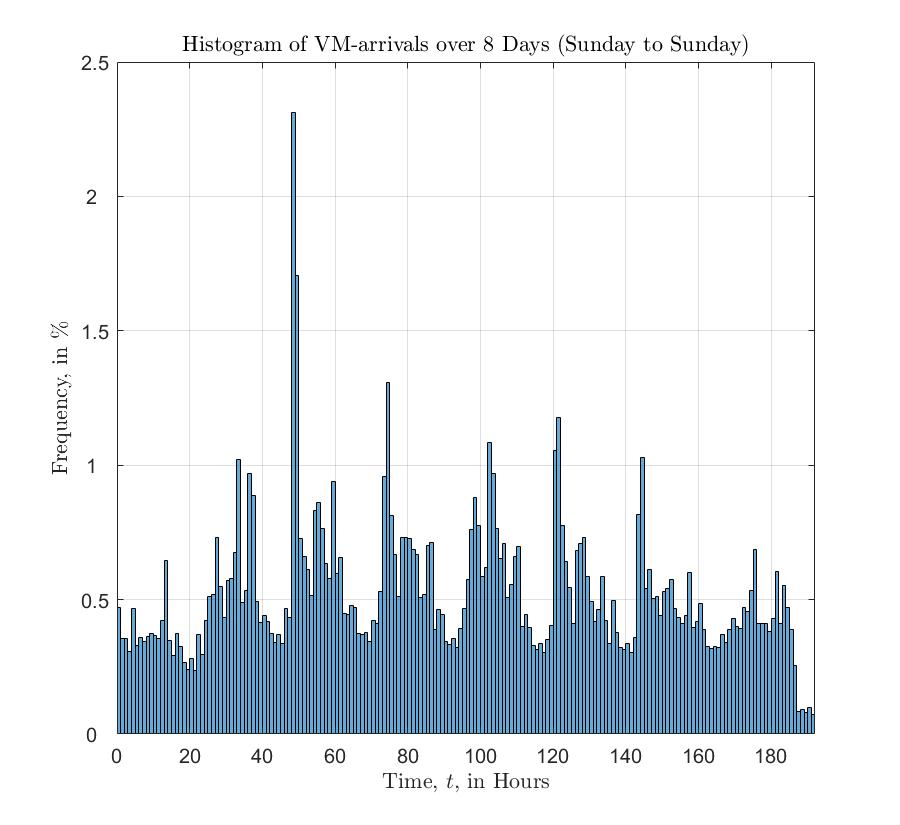}%
        \label{arrTimesH}%
         }%
     \hfill%
     \subfloat[]{%
        \includegraphics[width=0.53\textwidth]{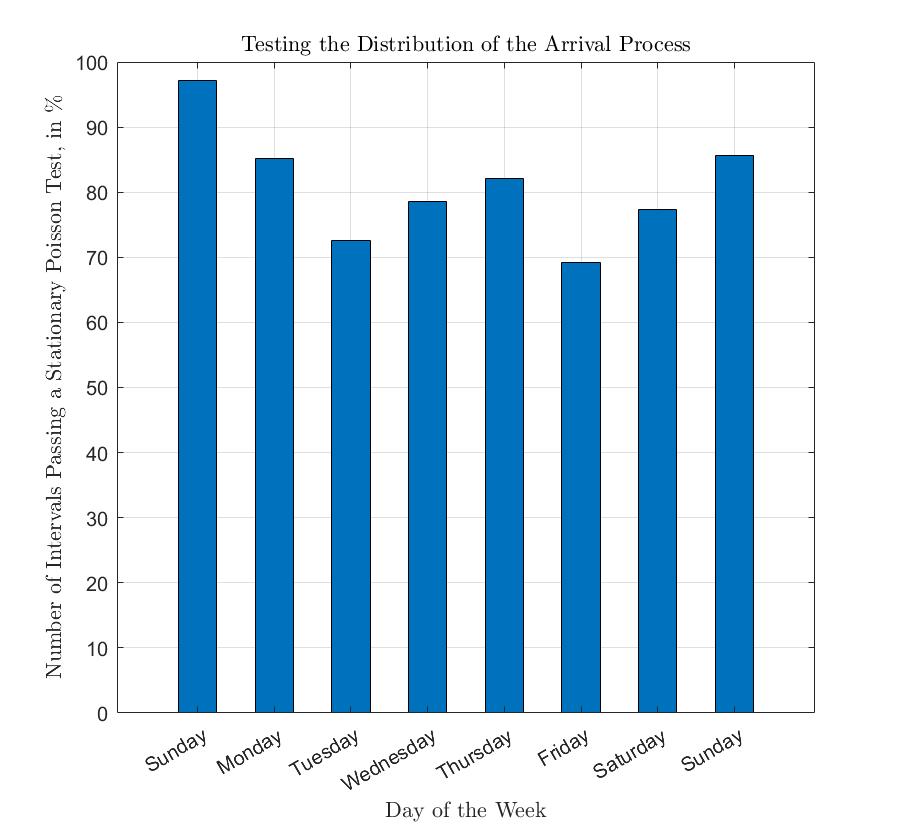}%
        \label{poisTest}
         }
\end{figure}

In Figure~\ref{arrTimesH}, we express counts of arriving to the system VMs as a percentage of the total demand recorded in the data set over the planning horizon (i.e., 8 days or 192 hours). According to this figure, the typical intensity of arrivals to the cloud computing servers follows a periodic quasiconcave pattern on each day from Sunday to Sunday, i.e., it reaches a maximum in every 24-hour period. Further, the magnitude of the maxima is consistent from Monday to Saturday with fewer VMs recorded on Sundays and a relatively taller spike in demand observed on Tuesday. We also notice that the arrival count rarely reduces below a base-level, i.e., approximately 0.3\% of the total demand, and it typically exceeds this quantity fourfold at the maxima over the week days. 

We follow the procedure described in \citet{brown2005statistical}
to test the null hypothesis that batch arrivals to the cloud computing servers follow a Poisson distribution with a piecewise constant rate. To this end, we break up the planning horizon into non-overlapping time intervals (approximately 5 minutes) and test the hypothesis of stationarity by applying the
Kolmogorov-Smirnov (KS) test and confirming that, for each time interval, the logarithmically transformed arrival times can be modeled by independent standard exponential
random variables. In Figure~\ref{poisTest}, we show that the 5-minute intervals over individual days typically pass the test in 80\% of instances. For example, the logarithmically transformed arrival times in nearly 85\% of 5-minute intervals on Monday can be modeled by independent standard exponential random variables. 
\subsection{Modelling Implications and Choices}\label{sec:EDAconc}
Our exploratory data analysis shows that distributions of service requirements of individual jobs in our data set typically have a thin tail, i.e., a majority of VMs request only a few CPU cores or units of RAM and remain in service for less than 5 hours. Further, jobs arrive in small batches with an average size of approximately 1.59. This reduces the effect of outliers on the expected demand estimated from the data set. Because the standard deviation of service durations within each batch tends to be smaller than 1 minute, we believe that complex modelling solutions accounting for an arbitrary variability in service times within a batch of jobs are not justified. To avoid unnecessary complexity, we build our analysis under the assumption that service times of jobs belonging to the same batch are identical. To deal with few instances of batches that have larger variability in respective service durations, we split them into several groups of jobs.     

According to Figure~\ref{arrTimesH}, arrivals to our system follow a time-varying and periodic pattern (with a period of 1 day). We describe the distribution of demand in Section~\ref{sec:arrPro}. Because jobs arrive for processing in relatively consistent quantities across weekdays, we choose a typical 24-hour period as a planning horizon for our modelling approach. To this end, we assume that our system has been in continuous operation for a long time, i.e., it is ergodic. Finally, 
our implementation of the statistical test set in \cite{brown2005statistical} demonstrates that batch arrivals to Huawei Cloud's servers on a typical weekday can be modelled by a non-homogeneous Poisson process reasonably well. According to Figure~\ref{poisTest}, the average test statistic exceeds 80\%. Thus, modelling demand by a compound non-homogeneous Poisson process is an appropriate choice. 
\section{Literature Review and Contributions}\label{Sec:LitRev}
Our paper introduces a stochastic model for a parallel processing system with heterogeneous workload and multiple types of resources, one of which is assumed to be dominant (i.e., most expensive or constraining). We modify this notion of a dominant resource borrowed from the communications literature \citep[see][for instance]{ghodsi2011dominant,zhang2013heterogeneity,wang2014dominant,li2017further}, which links it to proportions of the total demand in a system with multiple resources. 
In our system, jobs simultaneously occupy multiple units of resources and release them at service completion. Similarly to \cite{green1981comparing}, our queueing discipline complements the FIFO (first-in-first-out) service policy with utilization maximization to avoid service starvation (see \cite{bramson1994instability,gurvich2018collaboration, zychlinski2020managing} for more details). 

Pioneering work on queues with parallel processing is presented in~\cite{gimpelson1965analysis,wolman1972camp}. The former models a facility carrying two types of calls that require multiple transmission channels, while the latter analyzes a system where messages wait in parallel queues until they are transmitted to all receivers. Seminal applications of queues with parallel processing are discussed in \cite{green1978queues,green1980comparing,kaufman1981blocking}. Further, \cite{reiman1991critically} studies a multi-class loss system with parallel processing and derives its blocking probabilities. Parallel processing systems with an infinite buffer are considered in \cite{bambos1993scheduling}. Several researchers refer to queues with parallel processing as simultaneous resource possession systems \citep[see][for instance]{jacobson1982analyzing,kaufman1989conservation,kaufman1990approximate,visschers1999product}. 
In general, 
deriving analytical results for parallel processing queues is challenging \citep{papadimitriou1994complexity}. Thus, 
stochastic simulations \citep{feldman2010using,ma2016efficient,zychlinski2020managing,zychlinski2020scheduling} have been employed in recent years. 
We contribute to the literature by proposing a diffusion approximation \citep[e.g.,][]{halfin1981heavy,yao1985refining,whitt2004diffusion} of a periodic time-varying offered load in a system with parallel processing and a dominant resource. Our analysis builds on the stochastic offered load (SOL) for an $M_t/G/\infty$ queue \citep{eick1993physics,whitt2013om}, which we employ for making subsequent capacity allocation decisions.  

Since customers can submit multiple jobs to be processed at the same time, our paper contributes to the literature on queues with batch arrivals \citep[see][for instance]{chaudhry1983first,chaudhry1999modelling,antonis1999infinite,afeche2014double}. We model arrivals to our system by a non-homogeneous compound Poisson process with a general distribution of batch size. 
While many authors investigate multi-class queueing systems with arrivals that follow a
non-homogeneous Poisson process (see the survey papers by
\cite{defraeye2016staffing} and \cite{whitt2016queues}), time-varying queues with batch arrivals remain less prevalent in the queueing literature.
\cite{fakinos1984infinite}, for instance, 
considers a queue with batch arrivals and infinite capacity and obtains the transient distribution of the queue length and of the departure process. For additional information on queues with batch arrivals refer, for instance, to \cite{chatterjee1989non,suhasini2014queueing,kumar2017single,hemanth2020compound}. We contribute to this stream of literature by developing a modelling approach that adds batch arrivals to the queues with parallel processing and remains tractable at the same time. We achieve this goal by assuming that jobs within each batch share identical service times, while service duration of jobs from different batches is generally distributed. We show that under such assumption, the workload in our system follows a non-homogeneous compound Poisson process (this result does not hold in general, see~\cite{antonis1999infinite}). In addition, while we fit a time-varying batch arrival rate over a typical weekday using a family of polynomial functions, our methodology accommodates other functional forms, such as an exponential decay \citep{furman2021customer}, polynomial splines \citep{anderson1995smoothing}, or the sum of generalized sine functions \citep{yom2014erlang}.

We employ the SOL to develop a three-stage capacity allocation procedure that determines the optimal non-stationary allocation of several types of resources to
multiple classes of workload while meeting a set of pre-specified service level agreements (SLAs). Thus, we also contribute to the literature on capacity planning for queueing systems with parallel processing, time-dependent arrivals with batches, and fully flexible servers. In general, several studies investigate stationary or near-stationary staffing of queueing
systems with time-dependent arrival processes; see \citealp{bassamboo2009data, defraeye2013controlling} and
\cite{niyirora2016optimal} for example. The problem of staffing for a dynamic
service system has also been an active area of research (e.g.,
\citealp{henderson1999heuristic,akcali2006network,gans2012parametric}). However, literature on capacity planning for systems with parallel processing is scarce. Several authors propose an integer program to manage different types of idleness in such systems \citep{gurvich2015collaboration,zychlinski2020scheduling}. This type of research is generally oblivious to the patterns in the arrival rate and assumes that batch arrivals are not allowed. Our capacity allocation approach builds on the infinite server approximation (ISA) by~\cite{feldman2008staffing} and the square-root staffing (SRS) rule \citep[see, e.g.,][]{janssen2011refining, liu2018staffing} that we apply over a planning horizon of an arbitrary length. To the best of our knowledge, this is the first approach
that incorporates batch arrivals and time-varying demand in systems with parallel processing while preserving tractability and allowing practitioners to make capacity allocation decisions in real time. 

In addition, we contribute to research on capacity allocation for technological services by
proposing a methodology that is motivated by the operations of 
Huawei Cloud, a major provider of cloud computing services around the world. Setting levels of cloud capacity in an environment with high volatility in demand has been studied by many researchers.   
For instance, \cite{jiang2012self}, \cite{dorsch2014combining}, and
\cite{de2016impact} propose real-time capacity allocation schemes that can be
dynamically adjusted depending on the utilization of servers. The objective of
these methods is to maximize revenue or minimize costs by solving a stochastic
dynamic program. Several studies model cloud computing operations as a queueing
system (for example, \citealp{khazaei2011performance,vilaplana2014queuing,chiang2014profit}). This literature focuses on classical queueing-theoretic
approaches that model cloud services as open networks. Further, \cite{FurmanDiamant2021} models a private cloud service as a queueing system with retrials, homogeneous workload and a single type of capacity. They propose a stationary optimal policy allocating CPU cores in order to meet a pre-specified SLA. \cite{bergsma2021generating} combine elements of queueing framework with recurrent neural networks to predict future demand of a cloud service provider. Other authors focus on establishing dynamic pricing policies to maximize
cloud's revenue (e.g., \citealp{wang2010distributed, jin2014towards, chen2019pricing}).    
Our work extends these studies by proposing a queueing-based methodology that incorporates more realistic assumptions (e.g., time-varying demand,  batch arrivals, and multiple types of resources) and equips cloud service operators with a computationally efficient tool allowing them not only to determine the optimal non-stationary capacity of CPU cores and basic units of memory but also to forecast future workload. 

\section{Model Formulation}\label{sec:modelForm}
Consider a parallel processing system $\mathcal{QN}$ with $I+1$ classes of jobs: virtual machines (class $0$) and $I$ classes of containers. Class-$0$ jobs leave unserved if capacity is not available. If a container-type job is not served immediately, it joins an infinite queue with a pooled structure and a priority discipline, i.e., a job at the head of the queue is admitted to service if enough resources are provisioned, otherwise, to avoid unnecessary idleness, following jobs in order of their arrival can be serviced. Abandonment is not allowed and service contains $N$ types of resources, e.g., CPU cores and memory.
We define index sets $\mathcal{I}$
and $\mathcal{N}$, such that $|\mathcal{I}|=I+1$ 
and $|\mathcal{N}|=N$. Then, the scarcest or most expensive resource $\tilde{n}\in\mathcal{N}$ (as determined by a cloud operator)  is considered dominant. 

Further, we assume that class-$i$ jobs, $i\in\mathcal{I}$, follow a compound non-homogeneous Poisson arrival process $\{\Gamma_i(t)|t\ge 0\}$, i.e., contain independent and identically distributed (i.i.d.) batch sizes with mean $v_i<\infty$ and standard deviation $\beta_i<\infty$.  
 Define $\boldsymbol{s}(t)\coloneqq(s_1(t),s_2(t),...,s_{N}(t))\in\mathbb{Z}^{N}_{+}$ to be the total capacity of each type provisioned in the system at time $t$. A class-$i$ job requires resources of all types in order to start service, and its type-$n\in\mathcal{N}$ resource requirements are i.i.d. random variables; they are generally distributed with mean $r_{in}<\infty$ and standard deviation $\delta_{in}<\infty$. 
Preemption is not allowed, and jobs release resources allocated to them at the end of their service. 
 
\begin{figure}
\centering
\caption{Dynamics of the Service System}
\label{Fig:dynamicsQN}
\begin{tikzpicture}
 \draw[draw=black,dashed,fill={yellow!20}] (-0.75,-1.5) rectangle ++(3.75,5.5);
  \path (-2.5,0.5) pic {queuei};
 \draw[draw=black,fill={gray!50}] (-0.5,1.5) rectangle ++(3.25,2);
  \node[rectangle split,rectangle split parts=2, draw, minimum width=.5cm,rectangle split part fill={red!60}]  at (0,2.3) {};
  \node at (0,3.2) {\scriptsize$1$}; 
  \node[rectangle split,rectangle split parts=1, draw, minimum width=.5cm,rectangle split part fill={blue!30}] at (0.75,2.3) {};
  \node at (0.75,3.2) {\scriptsize$2$}; 
  \node[rectangle split,rectangle split parts=2, draw, minimum width=.5cm,rectangle split part fill={blue!30}]                    at (2.25,2.3) {};
  \node at (1.5,2.3) {...}; 
  \node at (2.25,3.2) {\scriptsize$N$}; 
\draw[draw=black,fill={gray!50}] (-0.5,-1) rectangle ++(3.25,2);
  \node[rectangle split,rectangle split parts=1, draw, minimum width=.5cm,rectangle split part fill={red!60}] at (0,-0.2) {}; {};
   \node at (0,0.7) {\scriptsize$1$};
  \node[rectangle split,rectangle split parts=3, draw, minimum width=.5cm,rectangle split part fill={blue!30}] at (0.75,-0.2) {};
  \node at (0.75,0.7) {\scriptsize$2$};
  \node[rectangle split,rectangle split parts=2, draw, minimum width=.5cm,rectangle split part fill={blue!30}]                    at (2.25,-0.2) {};
  \node at (1.5,-0.2) {...};
  \node at (2.25,0.7) {\scriptsize$N$};
   
   
\node at (1,1.35) {\vdots};
      
  
\node at (-5,0) (dummy1) {};

\node at (-3,0) (dummy4) {};

\node at (-8,0) (dummy7) {};

\node at (2.9,0) (dummy8) {};

\node at (5.5,0) (dummy11) {};

\node at (-4.275,2.65) (dummy14) {};
\node at (-4.275,1.5) (dummy15) {};
\node at (-4.4,1.35) (dummy21) {};

\node at (2.9,2.5) (dummy16) {};
\node at (5.5,2.5) (dummy17) {};

\node at (-8,2.5) (dummy18) {};
\node at (-0.6,2.5) (dummy19) {};
\node at (-8,1.35) (dummy20) {};

\draw [] (dummy7.east) to node [left,above]{\scriptsize Containers, Class $1, 2,...,I$}(dummy4.east);

\draw [paths] (dummy8.east) to node [above]{\scriptsize Served Jobs}(dummy11);
\draw [paths] (dummy16.east) to node [above]{\scriptsize Served Jobs}(dummy17);

\draw [paths,-] (dummy14.south) to node []{}(dummy15.south);
\draw [paths,->] (dummy21.east) to node [above]{\scriptsize Unserved Jobs}(dummy20.east);

\draw [paths] (dummy18.east) to node [left,above]{\scriptsize Virtual Machines,  Class $0$}(dummy19.west);

\node[] at (1.1,4.25)[] {\scriptsize Total Capacity};

\end{tikzpicture}
\end{figure}
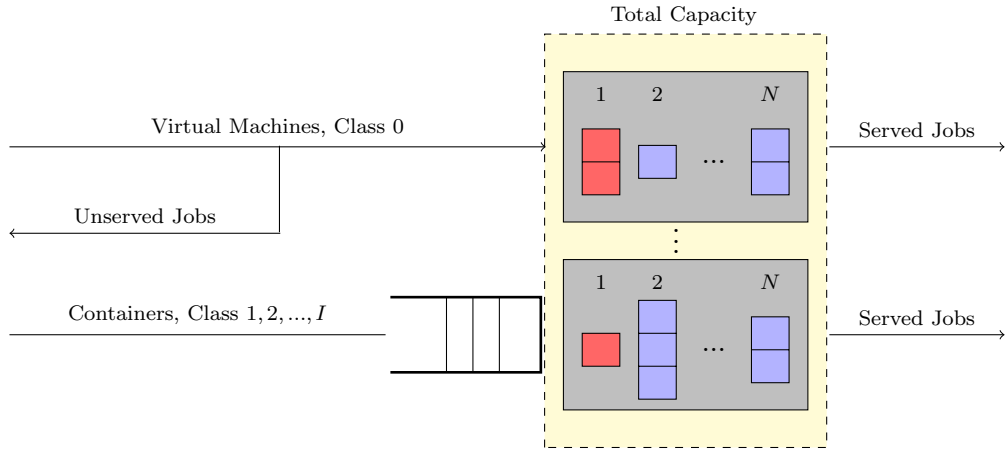

We denote the service duration of class-$i$ jobs from different batches by $S_i$, i.i.d. random variables with cdf $G_i$, mean $1/\mu_i<\infty$, and standard deviation $\sigma_i<\infty$. According to the data set provided to us by Huawei Cloud, class-$i$ jobs in the same batch have similar durations of service (see Section~\ref{sec:dscSum}). To ensure tractability, we assume that their durations are identical.
 Finally, let $\{X_{in}(t) \ | \ t\ge 0\}$ be 
 a stochastic process that describes the number of type-$n$ servers occupied by class-$i$ jobs at time $t$. 
 The workflow of such system is presented in Figure~\ref{Fig:dynamicsQN}. The yellow area in the figure represents service capacity, while the grey rectangles visualize capacity occupied by a single job. For example, the first job in Figure~\ref{Fig:dynamicsQN} requires two  type-1 resources, one type-2 resource, and 2 type-$N$ resources. Note that multiple jobs can be serviced at the same time. 
The red and blue squares represent units of dominant and non-dominant resources, respectively.  

Our partner organization would like to determine the minimum capacity $\boldsymbol{s}^*(t)$, such that a pre-specified service level agreement (SLA) is satisfied. Denote the stochastic process  measuring the time to access service (i.e., the service delay) experienced 
by class-$i$ jobs arriving to the system at time $t$ by $\{D_i(t,\boldsymbol{s}(t)) \ | \ t>0\}$. 
Then, on average, for any time $t$, $i\in\mathcal{I}$, $\alpha_i\in(0,1)$, and $\tau_i\in[0,\infty)$, no more than fraction $\alpha_i$ of type-$i$ jobs should experience a service delay longer than $\tau_i$ over a planning horizon $T>0$. 
 Due to independence of increments of a non-stationary compound Poisson process, the lack of anticipation assumption (LAA) holds \citep{wolff1982poisson,melamed1990arrivals}, and arrivals to the system observe time averages (ASTA). Thus, a fraction of jobs that wait for service longer than a pre-specified threshold $\tau_i$ can be represented by a proportion of time or probability that $D_i(t,\boldsymbol{s}(t))$ exceeds $\tau_i$. Then, we formulate the following optimization problem
\begin{align}\label{OPT}\tag{OPT}
\underset{\boldsymbol{s}(t)\in\mathbb{R}^{N}_{>0}}{\min} \boldsymbol{s}(t) & \ \textrm{
subject to } &\\
    &\mathbb{P}
    (D_i(t,\boldsymbol{s}(t))>\tau_i)\le \alpha_i 
    \quad \forall t\in[0,T], \quad\forall i\in\mathcal{I}.\label{def_SLA}\tag{SLA}&
\end{align}
The form of \ref{def_SLA} employed by \ref{OPT} has been studied in the call centre literature, and it is a preferred choice of modelling quality of service constraints \citep[see][for instance]{soh2017call}. Unfortunately, solving \ref{OPT} is challenging because it is a multi-dimensional, nonlinear minimization problem.
In addition, the probability distribution of the service delay in a queue with parallel processing is not tractable in closed form. To overcome these difficulties, in the following sections, we introduce a capacity allocation approach that constructs a heuristic solution to \ref{OPT} in three stages. Specifically, in stage 1 ($\infty$-stage), we determine the probability distribution of the stochastic offered load (SOL) and propose its approximation. In stage 2 ($\alpha$-stage),
we employ our workload approximation to set capacity of the dominant resource independently of other resource requirements.
Finally, in stage 3 ($\epsilon$-stage), we assign capacity levels to non-dominant resources. 
\section{The Capacity Allocation Procedure}\label{sec:capAprTheory}
In the first stage ($\infty$-stage) of our procedure, we consider the dynamics of $\mathcal{QN}$ in an uncapacitated regime to derive its stochastic offered load (SOL). Deriving the SOL involves several challenges, as jobs arrive in batches and require multiple units of resources at the same time. Thus, we decompose $\mathcal{QN}$ into queueing networks with tractable dynamics. Consider a pair of systems $\mathcal{QN}^e$ and $\mathcal{QN}^{\mathbbm{1}e}$ that are embedded in $\mathcal{QN}$. Arrival times of $\mathcal{QN}^e$ represent batch arrival times of $\mathcal{QN}$, i.e., an individual job arriving to $\mathcal{QN}^e$ is a batch of jobs observed in $\mathcal{QN}$. In addition, as opposed to $\mathcal{QN}^e$, all jobs in $\mathcal{QN}^{\mathbbm{1}e}$ require a single unit of type-$n$ capacity. Then, type-$i$ jobs arriving to these auxiliary systems follow a non-homogeneous Poisson process $\{\Lambda_i(t) \ | \ t>0\}$ with rate $\lambda_i(t)\in\mathbb{C}^1$.

Denote $\mathcal{QN}$ with infinite capacity by  $\mathcal{QN}_{\infty}$. In such system, jobs do not interfere with each other, thus, $\mathcal{QN}^e_\infty$ and $\mathcal{QN}^{\mathbbm{1}e}_{\infty}$ can be decomposed into $I+1$ independent components.
The $i^{th}$ component of $\mathcal{QN}^{\mathbbm{1}e}_{\infty}$ is an independent $M_t/G/\infty$ queue.   
Let $\{N_{i}(t) \ | \ t\ge 0\}$ be a stochastic process describing a type-$n$ SOL generated by the $i^{th}$ component of $\mathcal{QN}^{\mathbbm{1}e}_{\infty}$. Notice that, for any $n\in\mathcal{N}$, the SOL is represented by the same stochastic process. Further, for any $i\in\mathcal{I}$, $N_{i}(t)$ is well studied in the literature -- it is a non-homogeneous Poisson process with mean $m_{i}(t)$ \citep{eick1993physics}
\begin{equation}\label{Nt_mean}
    m_{i}(t) = \int_{0}^{\infty}\left(\int_{t-\tau}^{t}\lambda_i(u)du\right)dG_i(\tau) \approx
    \lambda_i(t)\mathbb{E}[S_i].
\end{equation}
In the following Lemma, we leverage this result to determine the probability distribution of a type-$n$ SOL of $\mathcal{QN}$ generated by class-$i$ jobs. We also obtain its first two moments in closed form.
\begin{lemma}[Stochastic Offered Load]\label{lemma:offeredLoad}
Suppose $X^{\infty}_{in}(t)$ denotes utilization of type-$n$ capacity by class-$i$ jobs in $\mathcal{QN}_{\infty}$. Then, $X^{\infty}_{in}(t)$ is a compound non-homogeneous Poisson process. Further,
\begin{equation}\label{Xpn:moments}
\mathbb{E}[X^{\infty}_{in}(t)] = v_ir_{in}m_i(t),
\quad
Var[X^{\infty}_{in}(t)] = m_i(t)\left(\delta_{in}^2v_i + r_{in}^2\beta_i^2+v_i^2r_{in}^2\right).
\end{equation}
\end{lemma}
In general, a compound non-homogeneous Poisson process exhibits mathematically convenient properties. However, because it is defined as a stochastic sum of random quantities, it is difficult to specify (or simulate) without making additional assumptions. 
Thus, in the following proposition, by applying Lemma~\ref{lemma:offeredLoad} and by scaling up the arrival process in $\mathcal{QN}_{\infty}$ by factor $\xi$, we derive a diffusion approximation of $X^{\infty}_{in}(t)$ that is easy to use and is not burdened by such limitations.
\begin{proposition}[Diffusion Approximation of the SOL]\label{Pro:diffusion}
Suppose, for any fixed pair $i\in\mathcal{I}$ and $n\in\mathcal{N}$, $W_{in}(t)$ is the standard Wiener process, then, for $t>0$, the asymptotic relationship holds
\begin{equation}\label{wiener:approx}
X^{\infty}_{in}(t)\simeq \hat{X}^{\infty}_{in}(t)=X^{\infty}_{in}(0) - \mathbb{E}[X^{\infty}_{in}(0)]+ \mathbb{E}[X^{\infty}_{in}(t)]
+ \sqrt{Var[X^{\infty}_{in}(t)]}W_{in}(t).
\end{equation}
\end{proposition}
To complete the $\infty$-stage of our approach, we specify the probability distribution of $\hat{X}^{\infty}_{in}(t)$ and the equation of its $\gamma$-percentile path. To prove these results, we adapt the square-root staffing rule. 
\begin{corollary}\label{co:bounds}
Denote the CDF of a standard normal random variable by $\Phi$. Then, for $t>0$,
\begin{enumerate}
    \item $\hat{X}^{\infty}_{in}(t)$ is normally distributed; its expectation and variance admit closed-form expressions 
    \[
    \mathbb{E}[\hat{X}^{\infty}_{in}(t)]=X^{\infty}_{in}(0) - \mathbb{E}[X^{\infty}_{in}(0)]+ \mathbb{E}[X^{\infty}_{in}(t)],\quad Var[\hat{X}^{\infty}_{in}(t)]=tVar[X^{\infty}_{in}(t)].
    \]
    \item For $\alpha\in[0,1]$, the $\alpha$-percentile path of $\hat{X}^{\infty}_{in}(t)$ follows the equation
    \begin{equation}\label{prcAlpha}
        x_{in}^{\alpha}(t) = X^{\infty}_{in}(0) - \mathbb{E}[X^{\infty}_{in}(0)]+ \mathbb{E}[X^{\infty}_{in}(t)]+ \sqrt{tVar[X^{\infty}_{in}(t)]}\Phi^{-1}(\alpha).
    \end{equation}
\end{enumerate}
\end{corollary}
In the following section, 
we present the $\alpha$-stage of our approach, where we 
apply the SOL approximation to determine capacity of the dominant resource required to satisfy a pre-specified SLA. 
\subsection{The $\alpha$-Stage of the Capacity Allocation}\label{sec:capacity}
In the $\alpha$-stage of our approach, we determine capacity of the dominant resource $\tilde{n}\in\mathcal{N}$
independently of other resource requirements given pre-specified SLAs. We achieve this goal by collapsing the entire system workload into a fictitious class. Then, we solve $I+1$ non-linear optimization problems with a single decision variable to determine capacity levels required for a job of such fictitious class to satisfy each of the pre-specified SLAs in the original problem. Finally, we construct a convex combination of these solutions weighted by the SOL generated by respective job classes.

As mentioned in Section~\ref{sec:modelForm}, solving \ref{OPT} necessitates the closed-form expression of
$D_{i}(t,\boldsymbol{s}(t))$ 
that is difficult to obtain. To this end,
we show the relationship between the SOL and the delay process. 
Notice that $\sum_{i=0}^I\hat{X}_{i\tilde{n}}^{\infty}(t)-s_{\tilde{n}}(t)$ quantifies the aggregated amongst job classes workload waiting for the dominant resource to get serviced at time $t$.
Then, following \cite{kim2013estimating}, we employ the instantaneous Little's Law to approximate the delay experienced by an ``average'' job in the system, should all jobs be collapsed into a single fictitious class
\begin{equation}\label{littleLaw}
D_{\cdot\tilde{n}}(t,s_{\tilde{n}}(t))=\left(\sum_{i=0}^I\hat{X}_{i\tilde{n}}^{\infty}(t)-s_{\tilde{n}}(t)\right)/\sum_{i=0}^Iv_ir_{i\tilde{n}}\lambda_i(t). 
\end{equation}
Then, by substituting $s_{\tilde{n}}(t)$ with $\nu_{i\tilde{n}}(t)\in\mathbb{C}^1$ in \eqref{littleLaw},
i.e., the quantity of the dominant resource required by jobs of the fictitious class to satisfy the SLA of class $i\in\mathcal{I}$, we re-write \ref{OPT} as follows
\begin{align}\label{alphaOPT}\tag{OPT$_{\alpha}$}
\underset{\nu_{i\tilde{n}}(t)\in\mathbb{R}_{\ge 0}}{\min} \nu_{i\tilde{n}}(t) & \ \textrm{
subject to } &\\
    &\mathbb{P}
    (D_{\cdot\tilde{n}}(t,\nu_{i\tilde{n}}(t))>\tau_i)\le \alpha_i. 
    \quad \forall t,\label{def_alphaSLA}\tag{SLA$_{\alpha}$}&
\end{align}
As opposed to the original optimization problem, \ref{alphaOPT} 
is a nonlinear problem with a single decision variable, thus, its solution can be obtained in closed form. We would like to solve \ref{alphaOPT} for all $i\in\mathcal{I}$ and to weight its solutions by the offered load of respective job classes. In the following result, we generalize Corollary~\ref{co:bounds}, to quantify the aggregated workload in the system.  
\begin{corollary}\label{co:bounds_pooled}
For $t>0$,
    $\sum_{i=0}^I\hat{X}^{\infty}_{in}(t)$ is normally distributed. Further, we have that
    \[
    \mathbb{E}\left[\sum_{i=0}^I\hat{X}^{\infty}_{in}(t)\right]=\sum_{i=0}^I\left(X^{\infty}_{in}(0) - \mathbb{E}[X^{\infty}_{in}(0)]+ \mathbb{E}[X^{\infty}_{in}(t)]\right),\quad Var\left[\sum_{i=0}^I\hat{X}^{\infty}_{in}(t)\right]=\sum_{i=0}^ItVar[X^{\infty}_{in}(t)].
    \]
\end{corollary}
Similarly to Corollary~\ref{co:bounds}, we denote an $\alpha$-percentile path of the aggregated SOL by $x_{\cdot{}n}^{\alpha_i}(t)$. In the following proposition, we combine the square-root staffing rule with the aggregated SOL approximation to solve \ref{alphaOPT} in closed form. Notice that solutions to \ref{alphaOPT} are only optimal with respect to a specific choice of the SLA corresponding to class $i$. Thus, we call these solutions fictitious. 
\begin{proposition}[Fictitious Quantity of the Dominant Resource]\label{pro:Capacity} Define $\gamma_i\coloneqq1-\alpha_i$. Then, by Proposition~\ref{Pro:diffusion} and Corollary~\ref{co:bounds_pooled}, 
the solution to \ref{alphaOPT} admits the following expression 
\begin{equation}\label{capStage1}
\hat{\nu}_{i\tilde{n}}(t) = 
x_{\cdot{}\tilde{n}}^{\gamma_i}(t)-\tau_i\sum_{j=0}^Iv_jr_{j\tilde{n}}\lambda_j(t), \quad i\in\mathcal{I}.
\end{equation}
\end{proposition}
To conclude the $\alpha$-stage of our approach, we define a time-varying heuristic that provisions a quantity of the dominant resource given a set of pre-specified SLAs as a convex combination of the fictitious solutions obtained in Proposition~\ref{pro:Capacity}. Define $\omega_i(t) \coloneqq \frac{X_{i\tilde{n}}^{\infty}(t)}{\sum_{i=0}^IX_{i\tilde{n}}^{\infty}(t)}$, then we have that
\begin{equation}\label{capAlphaSol}
\hat{s}_{\tilde{n}}(t) \coloneqq \sum_{i=0}^I\omega_i(t) \hat{\nu}_{i\tilde{n}}(t), \quad \forall t.    
\end{equation}
In the analysis so far, we have proposed a heuristic that sets the capacity of the dominant resource given pre-specified SLAs. However, our results do not prescribe capacity levels to the rest of the resources. In the $\epsilon$-stage of our capacity allocation procedure, we
determine $s_n(t)$ for any $n\ne\tilde{n}\in\mathcal{N}$.
\subsection{The $\epsilon$-Stage of the Capacity Allocation}\label{sec:heuSol}
In the last stage of our approach, we would like to set the service capacity of resource $n\ne\tilde{n}$ high enough to allow jobs to get serviced once a sufficient quantity of the dominant resource becomes available. For an arbitrary small $\epsilon>0$ and $n\ne\tilde{n}$, we consider
the following minimization problem
\begin{align}\label{epsilonOPT}\tag{OPT$_{\epsilon}$}
\underset{s_n(t)\in\mathbb{R}^{N}_{\ge0}}{\min} s_n(t) & \ \textrm{
subject to } &\\
    &\mathbb{P}\left(\sum_{i=0}^IX_{in}^{\infty}(t)>s_n(t)\bigg|\sum_{i=0}^IX_{i\tilde{n}}^{\infty}(t)\le \hat{s}_{\tilde{n}}(t)\right)\le \epsilon, &\forall n\ne \tilde{n},\forall t,\forall i.\label{def_epsSLA}\tag{SLA$_{\epsilon}$}
\end{align}
In the proposition below, we show that 
when the capacity of non-dominant resources is set according to the solution to \ref{epsilonOPT}, only the solution to \ref{alphaOPT} determines if the SLAs in \ref{OPT} are satisfied.
\begin{proposition}[Setting Quantities of Non-dominant Resources]\label{pro:secRes}
Suppose that $\hat{s}_{\tilde{n}}(t)$ satisfies \ref{def_SLA} at any time $t\in[0,T]$ and $\alpha\coloneqq1-\epsilon$. 
Then, for the solutions to \ref{epsilonOPT}, such that $n\ne\tilde{n}$,
$\boldsymbol{\hat{s}}(t)\coloneqq(\hat{s}_{1}(t),\hat{s}_{2}(t),...,\hat{s}_{N}(t))$ is the solution to \ref{OPT}. Further, $\hat{s}_{n}(t)$ admits the following expression 
\begin{equation}\label{secCapExp}
\hat{s}_{n}(t) = x_{\cdot{}n}^{\alpha}(t), \quad i\in\mathcal{I}, \quad \forall n\ne \tilde{n}\in\mathcal{N}.
\end{equation}
\end{proposition}
Proposition~\ref{pro:secRes} sets 
the capacity levels of non-dominant resources so that they do not hinder servicing of jobs which have observed a sufficient quantity of the dominant resource specified in the $\alpha$-stage of the procedure. Overall, by assigning capacity to the dominant and non-dominant resources, our approach produces a heuristic solution to \ref{OPT} that aggregates workload in the system and results in a ``pooled'' policy. In the following section, we discuss an alternative procedure that is closer to the current practices of Huawei Cloud and is driven by aggregating resources that are dedicated to individual job classes. 
In our numerical study, we employ this policy as a benchmark.  
\subsection{Benchmark Approach: Current Practices at the Partner Organization}\label{sec:SiloPolicy}
As mentioned in Section~\ref{Sec:intro}, offerings of containerized services have become a trendy feature amongst cloud computing providers. Combining them with traditional virtual machines on an integrated platform is even more desirable. Huawei Cloud has been offering computing virtualization software to support VM functionality (Elastic Cloud Server) as well as a dedicated cloud container engine, i.e., their Kubernetes service, to clients on demand. 

The objective, however, is 
to ensure seamless availability of resources to a wide range of the company's offerings powered by VMs and containers alike, without a necessity to dedicate resources to a specific class of jobs~\citep{ke2021fundy}.

One way to compute the total capacity available in a system with resources that are dedicated to job classes is to 
combine them in a single pool. As opposed to our pooled policy, this approach (the benchmark policy) aggregates dedicated resources instead of doing so with respective workloads prior to making any capacity allocation decision. Define a dedicated to class-$i$ capacity of type-$n$ by $\pi_{in}(t)$, such that $\pi_n(t)\coloneqq\sum_{i=0}^I \pi_{in}(t)$. Then, for a resource $n\in\mathcal{N}$ 
we rewrite \eqref{littleLaw}
\begin{equation}\label{littleLawSilo}
D_{in}(t,\pi_{in}(t))=\left(\hat{X}_{in}^{\infty}(t)-\pi_{in}(t)\right)/v_ir_{in}\lambda_i(t). 
\end{equation}
In the following Corollary, we determine capacity levels $\boldsymbol{\pi}(t)\coloneqq(\pi_{1}(t),\pi_{2}(t),...,\pi_{n}(t))$ provisioned in the system such that pre-specified SLAs are satisfied for a class-$i\in\mathcal{I}$ and time $t\in[0,T]$.
\begin{corollary}\label{co:siloPol}
For $\gamma_i \coloneqq 1-\alpha_i$ and $i\in\mathcal{I}$, using Proposition~\ref{pro:Capacity} and Corollary~\ref{co:bounds}, we have that
\begin{enumerate}
    \item The optimal capacity of the dominant resource admits the following expression
    \begin{equation}\label{prcAlphaSilo}
        \hat{\pi}_{i\tilde{n}}^{\gamma_i}(t) = x_{i\tilde{n}}^{\gamma_i}(t)-\tau_iv_ir_{i\tilde{n}}\lambda_i(t),\quad \hat{\pi}_{\tilde{n}}(t) \coloneqq \sum_{i=0}^I \hat{\pi}_{i\tilde{n}}^{\gamma_i}(t).
    \end{equation}
    \item For $\gamma \coloneqq 1-\epsilon$, the optimal capacity of a non-dominant resource $n\ne\tilde{n}\in\mathcal{N}$ follows the equations
    \begin{equation}\label{prcEpsSilo}
        \hat{\pi}_{in}^{\gamma}(t) = x_{in}^{\gamma}(t),\quad \hat{\pi}_{n}(t) \coloneqq \sum_{i=0}^I \hat{\pi}_{in}^{\gamma}(t).
    \end{equation}
    \end{enumerate}
\end{corollary}
In a setting where capacity is reserved for job classes, equations \eqref{prcAlphaSilo} and \eqref{prcEpsSilo} specify 
optimal quantities of dominant and non-dominant resources, respectively. We estimate the total capacity of type-$n$ by aggregating its quantity amongst all job classes. In the following section, we discuss the modelling assumptions of the proposed pooled policy as well as the suggested benchmark. 
\subsection{Model Discussion}\label{sec:modelDisc}
Our pooled policy enables practitioners to determine quantities of multiple resources in accordance with pre-specified SLAs in a time-varying setting with heterogeneous demand and batch arrivals. Capacity allocation in systems with multiple resources is typically challenging and often employs a single resource abstraction, i.e., bundles of resources (or slots). 
However, in a realistic setting with
heterogeneous service requirements, 
this abstraction is 
inefficient~\citep{parkes2015beyond}. Instead, we borrow the notion of the dominant resource from the Dominant Resource Fairness (DRF) algorithm proposed in~\cite{ghodsi2011dominant} 
and modify it to denote the most constraining resource.

According to the DRF, a resource with the maximum proportion of respective demand to the available in the system capacity is dominant. Further, the DRF framework 
assumes a fixed number of customers who submit jobs with deterministic service requirements. 
At the end, to determine the optimal assignment of existing capacity to clients, the DRF procedure solves a linear program that maximizes minimum shares of dominant resources available to each customer. Contrary to the DRF, our model focuses solely on individual jobs that belong to the same or different batches. As the workload in our system is time-varying and stochastic, determination of the dominant resource is
driven by the historical data and an expert opinion, i.e., we disentangle this notion from individual jobs and re-define it as an exogenous property of a cloud computing system. This allows us to set levels of flexible (rather than dedicated) capacity myopically and respect probabilistic SLAs. In addition, 
our approach requires that jobs belonging to the same batch have identical service durations. This assumption is required to preserve tractability, and, thus, our data exploration in Section~\ref{sec:EDA} has shown that service durations of such jobs are indeed similar. 

Further, the delay process in the constraints of \ref{alphaOPT} and \ref{epsilonOPT} is constructed by the application of the Little's Law to a time-varying environment. According to \cite{kim2013estimating}, this natural approximation of the waiting time may exhibit a significant bias in highly time-varying systems with long service durations. Their study shows that the bias is proportional to three separate quantities: variability of the waiting time distribution, the relative slope of the arrival
function, and the magnitude of the estimated waiting time. Further, given an adequate approximation of the system workload, under a time-dependent square-root-staffing rule,
the waiting time distribution remains near stationary even though the arrival rate is time varying.  By aggregating arrival rates across all job classes, our approach ensures greater uniformity of the arrival process (which in turn reduces variability of the waiting time distribution) and decreases the relative slope of the arrival rate.  
In addition, the magnitude of the waiting time is influenced by service quality, and, thus, our methodology is to perform better in a setting with stricter SLAs. On the contrary, approaches that dedicate capacity to individual job classes do not benefit from pooling arrivals, i.e., they are more vulnerable to the estimation bias of the delay process in a time-varying setting.

Finally, any prediction of system dynamics that uses first or second order techniques is sensitive to changes in initial conditions. In a time-varying setting, this sensitivity may become a concern. Because our pooled policy requires the total start-of-the-day quantities of class-$i$ jobs observed in the system, its accuracy is influenced by variability in workloads at the end of a 24-hour period. However, aggregating offered load at the $\alpha$ and $\epsilon$ stages of our procedure smoothens this effect especially if demand of a class that contributes the most work to the system is the least time varying. 
\section{Numerical Study}\label{sec:NumStu}
In this section, we apply our three-stage procedure to a hypothetical cloud service and compare its performance to a benchmark policy described in Corollary~\ref{co:siloPol}. We first show that, in the uncapacitated system parameterized by the data, our approximation of the stochastic offered load (SOL), as derived in Proposition~\ref{Pro:diffusion}, confines the utilization of CPU cores 
within the 80-percentile range of its realizations. Then, we determine the levels of capacity required to satisfy the pre-specified SLAs in two practical scenarios. We consider a setting with 4 classes of containers that exhibit near stationary demand and contrast it with an environment where such demand is highly time-varying. Our experiments suggest that the heuristic solution to \ref{OPT} (i.e., the pooled policy) assigns approximately 20\% fewer resources than the benchmark while achieving better service quality.

For both time-varying and near stationary configurations of our discrete event simulations, we fix $\alpha_0$ and $\tau_0$ to 0.01 and 0, respectively, because the probability of waiting of the VMs approaches 0 asymptotically. Further, for $i\ne 0$, we fix $\alpha_i$ to 0.2. This ensures that SLAs of containers only differ in their waiting thresholds $\tau_i$. We distinguish 4 classes of containers with different levels of tolerance for waiting: low, moderate, and high. In particular, two classes of containers with strict SLAs have relatively low waiting thresholds of 90 minutes ($\tau_1=\tau_2$), while the other two classes 
have their waiting thresholds set to 450 minutes ($\tau_3$) and 900 minutes ($\tau_4$), respectively. 

\begin{figure}[h] 
\caption{Fitted Batch Arrival Rate of VMs}
\label{arrTimesFit}
     \subfloat[]{%
        \includegraphics[width=0.53\textwidth]{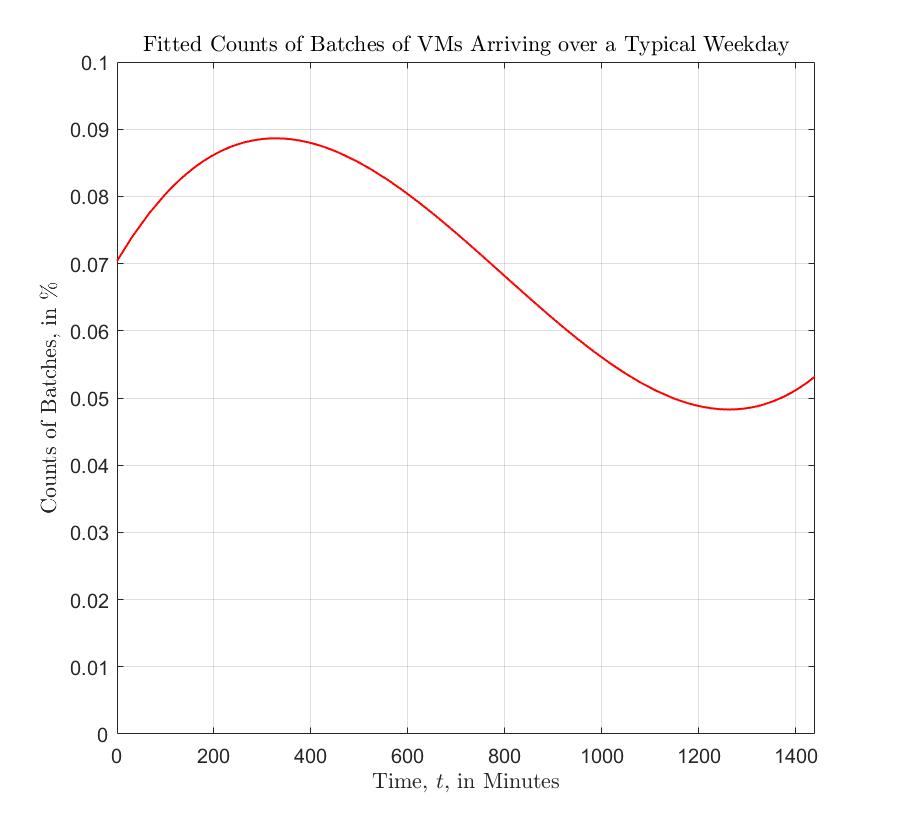}%
        \label{arrTimesFit1day}%
         }%
     \hfill%
     \subfloat[]{%
        \includegraphics[width=0.53\textwidth]{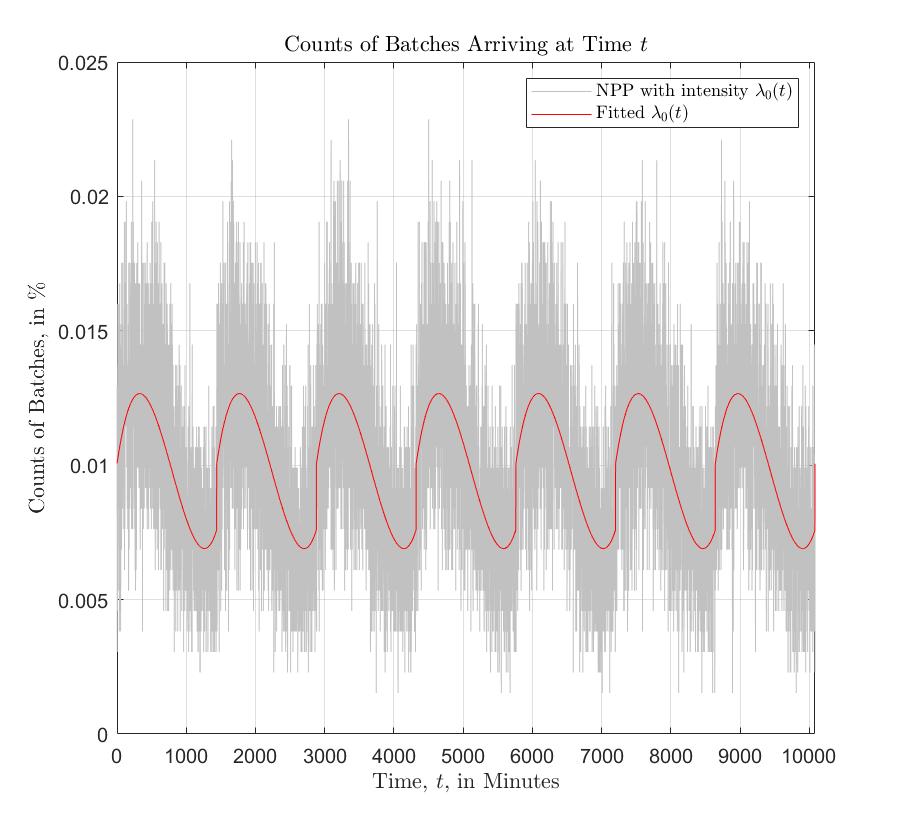}%
        \label{arrTimesFit8days}%
         }
\end{figure}

Following the argument in Section~\ref{sec:EDAconc}, since the arrival process is periodic and can be modelled by a non-homogeneous Poisson process, without loss of generality, we select
individual arrivals of the VMs over a weekday and
use the curve fitting toolbox in MatLab R2021a to fit the mean of the corresponding batch arrival process by a polynomial of the third order (see Figure~\ref{arrTimesFit1day}). For confidentiality, the counts of batch arrivals on the y-axis are displayed in percentage of the total quantity. We repeat this arrival pattern over the planning horizon of 7 days and generate batch arrivals with the fitted mean
using a standard thinning algorithm (see Figure~\ref{arrTimesFit8days}).

We model the batch arrivals of containers by independent non-homogeneous Poisson processes with their means defined as quadratic functions. As advised by the partner organization, we parameterize them such that the total daily demand of each class accounts for 
19,000 batches or 
approximately 29,000 individual jobs (see Appendix~\ref{appedA}). Figure~\ref{arrContRates} presents their batch arrival rates over a typical 24-hour period. Specifically, in the near-stationary experiments, $\lambda_1(t)$ is decreasing, $\lambda_2(t)$ and $\lambda_4(t)$ are increasing, and $\lambda_3(t)$ has a concave shape (see Figure~\ref{contNS}). In the time-varying setting, $\lambda_1(t)$, $\lambda_2(t)$, and $\lambda_4(t)$ are characterized by the spikes over two-hour intervals in the morning, afternoon, and evening, respectively (there are no arrivals otherwise), while $\lambda_3(t)$ has a spike of a lesser height, but remains positive over the period of 12 hours in the daytime (see Figure~\ref{contTV}). 
\begin{figure}[h] 
\caption{Batch Arrival Rates of Containers: (a) Near-stationary Setting; (b) Time-varying Setting.}
\label{arrContRates}
     \subfloat[]{%
        \includegraphics[width=0.53\textwidth]{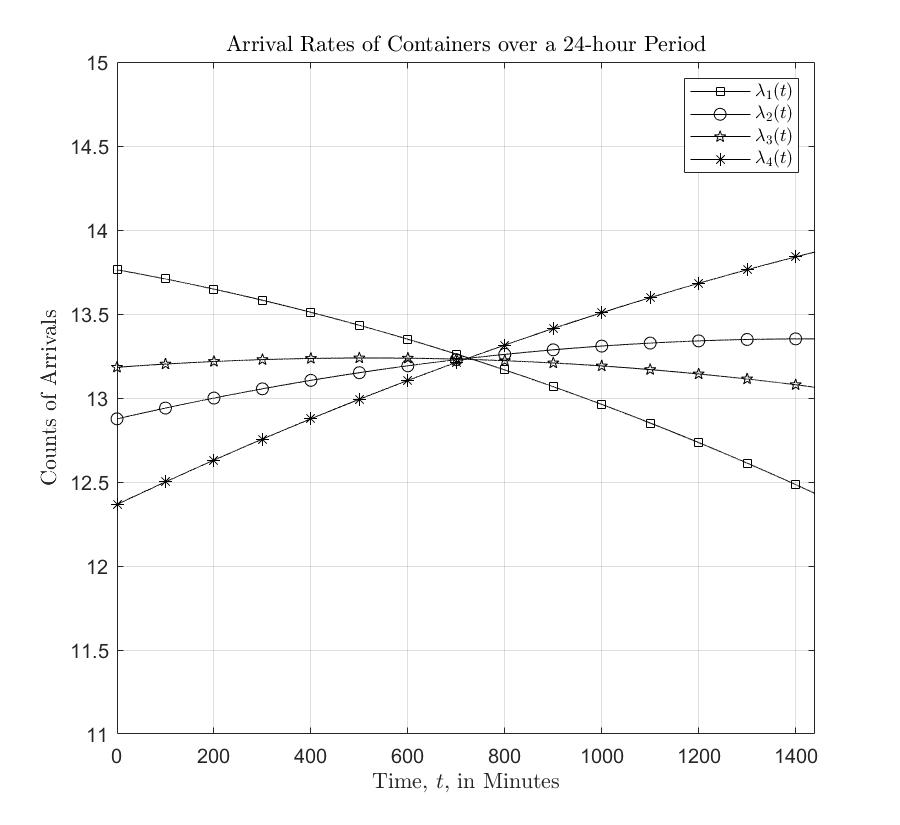}%
        \label{contNS}%
         }%
     \hfill%
     \subfloat[]{%
        \includegraphics[width=0.53\textwidth]{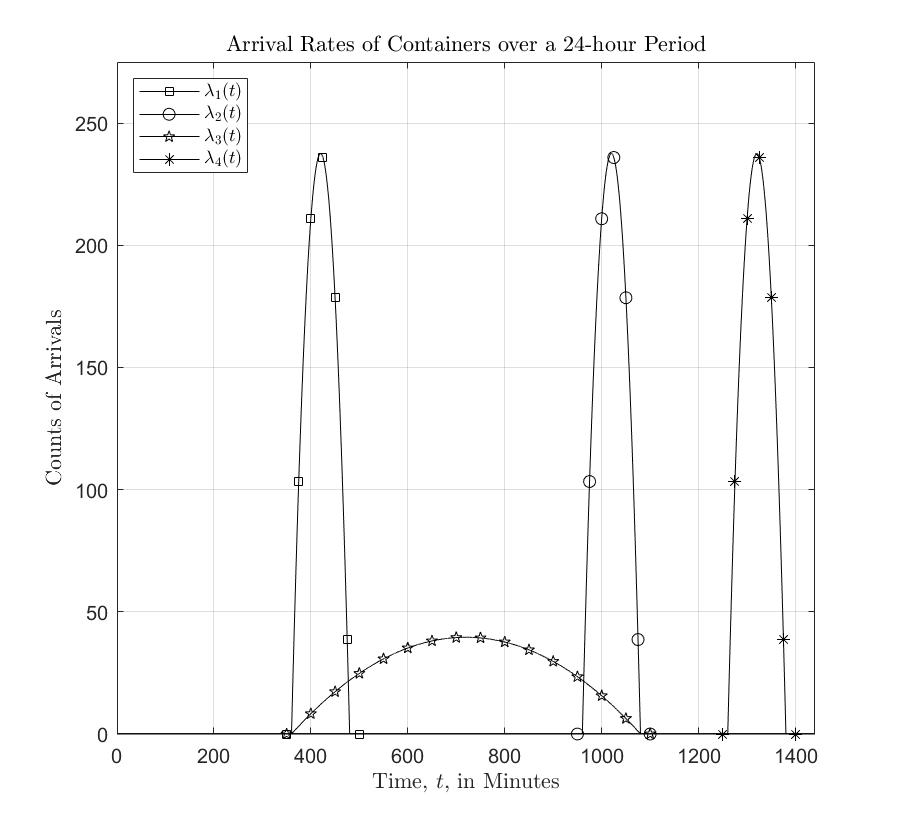}%
        \label{contTV}%
         }
\end{figure} 

We generate service durations of VMs and containers using a uniform sampling with replacement, i.e., we employ a bootstrap distribution of service times from the data set provided to us by Huawei Cloud. Because containers are relatively short-lived jobs as opposed to the VMs, we truncate their service time distributions, such that all realizations that exceed 8 hours are substituted by this maximum value. Further, we fit the probability mass functions (pmf) of the CPU-core and RAM requirements of VMs in the data set by employing a standard frequency-based technique. We assume that service requirements of VMs and containers are described by the same pmf and provide a summary of the parameters used in our discrete event simulations in Appendix~\ref{appedA}.    

We conduct simulations of the service system in Figure~\ref{Fig:dynamicsQN} operating under our heuristic (pooled) policy $\boldsymbol{s}(t)$, governed
by equations \eqref{capAlphaSol} and \eqref{secCapExp},
and 
the benchmark $\boldsymbol{\pi}(t)$
as described in Corollary~\ref{co:siloPol}. Because the latter reserves capacity for job classes, each experiment under such policy is equivalent to simulating systems with a single class $i$ for all $i\in\mathcal{I}$ (i.e., 5 simulations in total). We implement these time-varying policies by updating the total number of resources available in the system once per minute. Then, we perform 10 independent sets of these 6 experiments in the near-stationary and time-varying settings that account for 120 trials in total. At the end of each trial, we record the time-varying distribution of system utilization in percentage, the total number of VMs that leave the system unserved, 
and the times individual containers spend waiting to be serviced.   


\subsection{Fidelity of the SOL Approximation}\label{sec:SOLfid}
\begin{figure}[h] 
\caption{Data Driven CPU-core Workload: (a) 80-Percentile Range of Ranked Paths; (b) Percentile Translation.}
\label{solFid}
     \subfloat[]{%
        \includegraphics[width=0.53\textwidth]{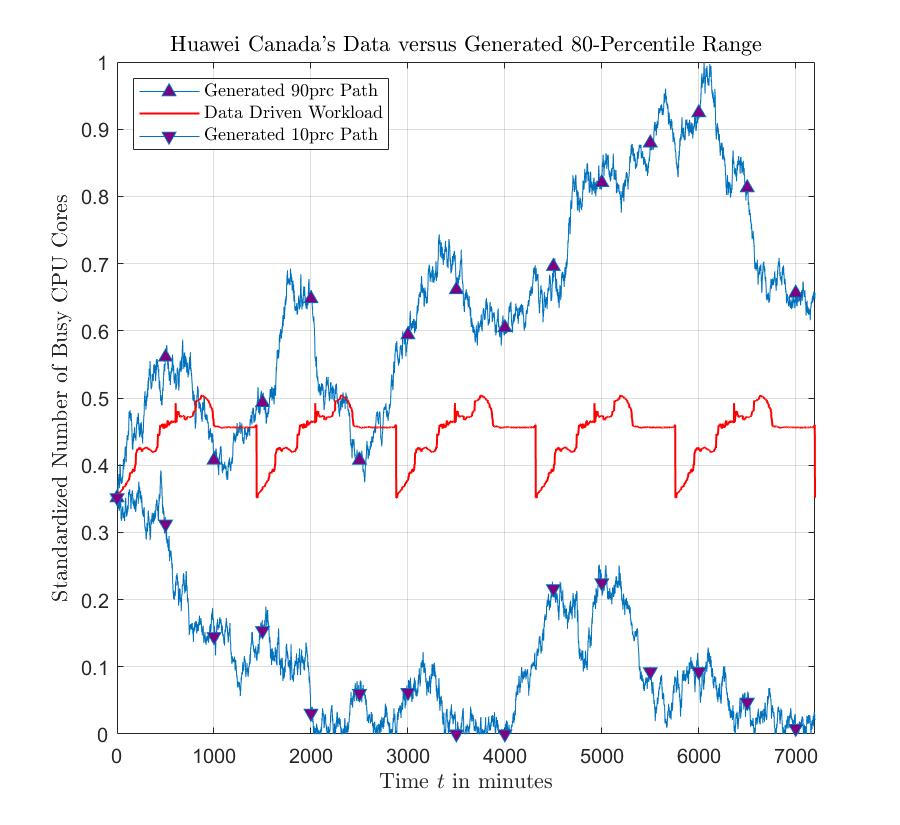}%
        \label{prcTable}%
         }%
     \hfill%
     \subfloat[]{%
        \includegraphics[width=0.53\textwidth]{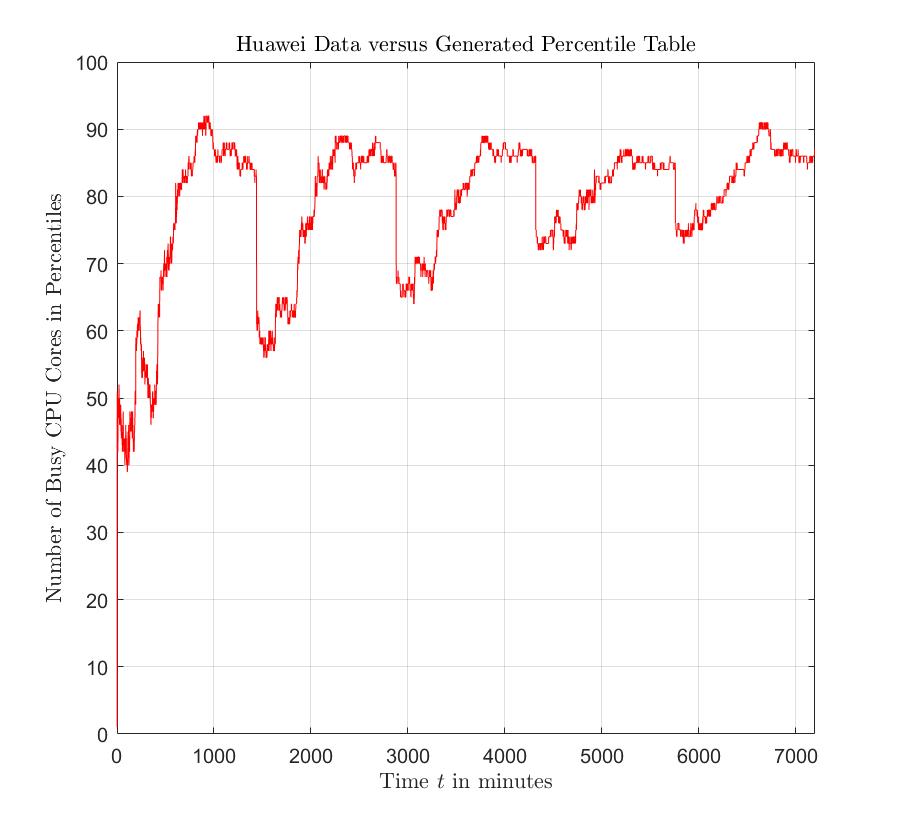}%
        \label{prcRange80}%
         }
\end{figure}

In this section, we confirm that our SOL approximation as derived in Proposition~\ref{Pro:diffusion} tracks the workload observed in the data set with relatively high fidelity. To this end, applying equation \eqref{wiener:approx} and Corollary~\ref{co:bounds} to the offered load of the dominant resource (CPU cores), we generate 200 stochastic paths over a planning horizon of 5 days. Then, we discretize it into 7200 one-minute intervals. We rank these paths, so that the realization with most arrivals over the largest number of one-minute intervals is labeled first, i.e., a 99.5-percentile path. According to Figure~\ref{prcTable}, the respective 10 and 90-percentile paths constructed by this ranking procedure confine the CPU-core workload observed in the Huawei Cloud's data set (see the red line in the figure).

In addition, using the same set of 200 stochastic paths, we construct their empirical distributions, i.e., at a given minute, the largest number in the array of 200 values is marked as a 100-percentile. Then, on the y-axis of Figure~\ref{prcRange80}, we translate the offered load of the CPU cores observed in the data set into the values of empirical percentiles. We find that workload implied by the data set oscillates between 70 and 90-percentile thresholds once it reaches a steady state.
\subsection{Discrete Event Simulation: Containers with Near Stationary Demand}\label{sec:simSta}
In this setting, 
the demand of containers slowly increases or decreases over a typical day, i.e., customers submit their requests for service with a near-uniform intensity over the planning horizon  (see Figure~\ref{contNS}). We first compare the capacity levels allocated by our pooled policy and the benchmark described in Section~\ref{sec:SiloPolicy}. We then investigate the system utilization under these policies and plot survival functions of waiting times for all classes of containers in order to verify that their SLAs are satisfied in the discrete event simulations. Our results suggest that, in the near-stationary setting, the proposed fully flexible heuristic dominates the benchmark by achieving better quality of service while assigning fewer units of capacity to the cloud computing system.

\begin{figure}[h] 
\caption{Near-stationary Setting: Dynamics of the Total Capacity Allocation over a Typical Day.}
\label{allocstat}
     \subfloat[]{%
        \includegraphics[width=0.53\textwidth]{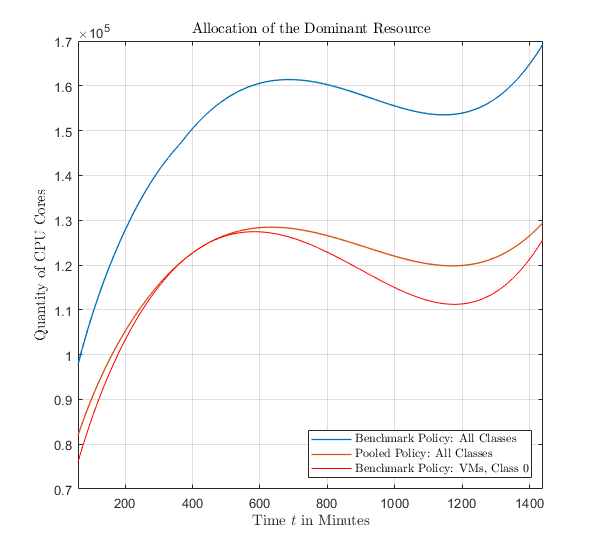}%
        \label{allocCPUstat}%
         }%
     \hfill%
     \subfloat[]{%
        \includegraphics[width=0.53\textwidth]{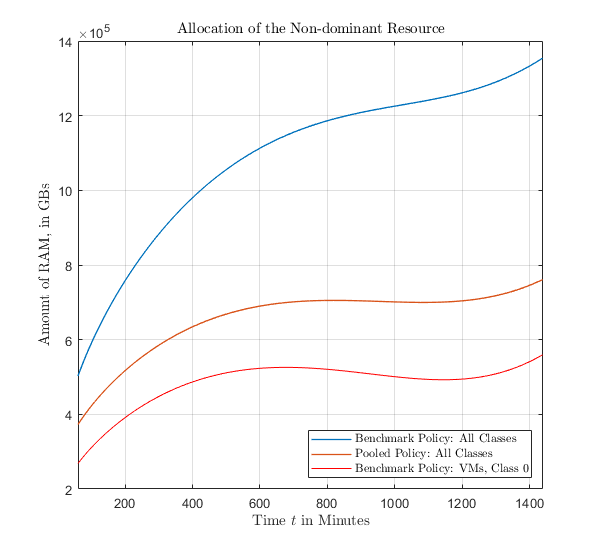}%
        \label{allocRAMstat}%
         }
\end{figure}

According to Figure~\ref{allocstat}, the combined capacity reserved by the benchmark contains at least 20\% and 50\% more CPU cores and units of memory, respectively, compared to the quantities assigned by our pooled policy. We notice that
the capacity allocated to VMs by the benchmark (see the red curve in the figure) imposes a lower bound on the total capacity levels. 

\begin{figure}[h] 
\caption{Near-stationary Setting: Distributions of the Capacity Utilization over a Typical Day.}
\label{utilstat}
     \subfloat[]{%
      \includegraphics[width=0.53\textwidth]{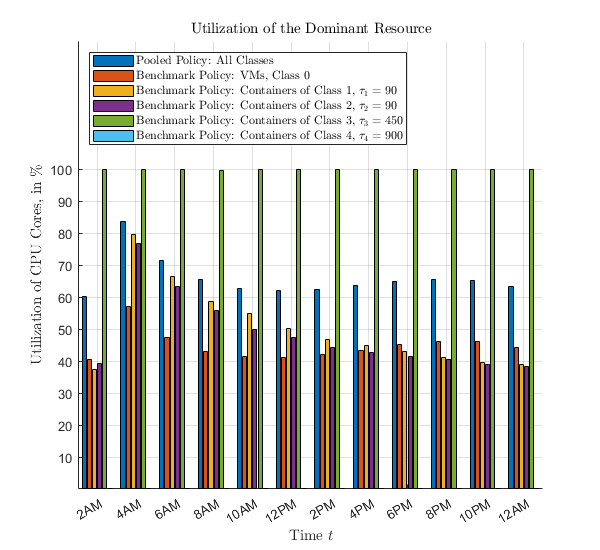}%
        \label{utilCPUallStat}%
         }%
     \hfill%
     \subfloat[]{%
        \includegraphics[width=0.53\textwidth]{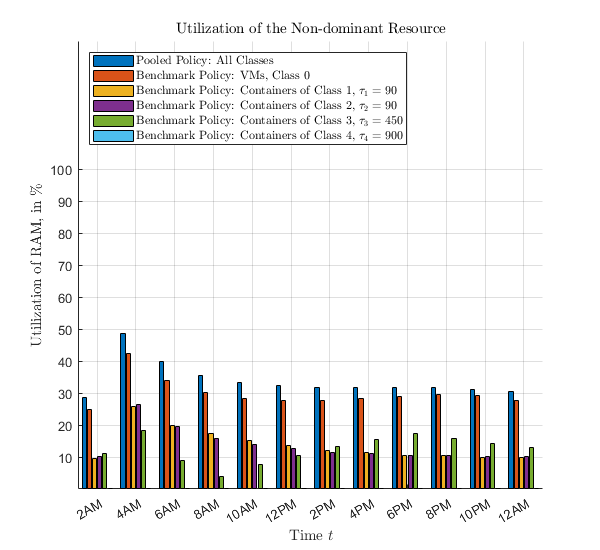}%
        \label{utilRAMallStat}%
         }
\end{figure}

Further, in Figure~\ref{utilCPUallStat}, we show that both policies may result in idling of resources. However, the utilization of CPU-cores under the pooled policy
rarely exceeds the 75\% mark and remains below 100\% at all times, i.e., no job has to wait for service.
In contrast, dedicating resources results in service congestion. While containers with stricter SLAs ($\tau_1=90$ and $\tau_2=90$) experience idling of resources, jobs with a higher waiting threshold ($\tau_4=900$) fail to get serviced due to insufficient quantity of CPU cores. Notice that the capacity reserved for these jobs remains idle as no CPU core is provisioned to them. 
According to Figure~\ref{utilRAMallStat}, the non-dominant resource does not hinder servicing of jobs under any of the policies. 

\begin{figure}[h] 
\caption{Near-Stationary Setting: (a) Likelihood of  Delays; (b) Counts of Delayed Jobs.}
\label{waitstat}
      \subfloat[]{%
     \includegraphics[width=0.53\textwidth]{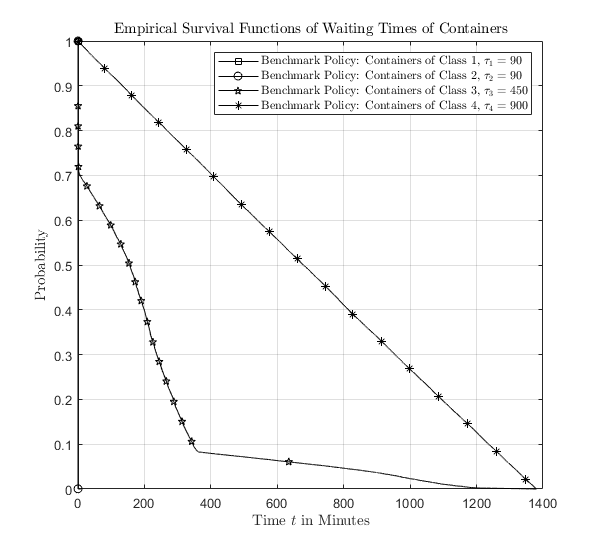}%
          \label{survWaitstat}%
          }%
      \hfill%
      \subfloat[]{%
       \centering \includegraphics[width=0.53\textwidth]{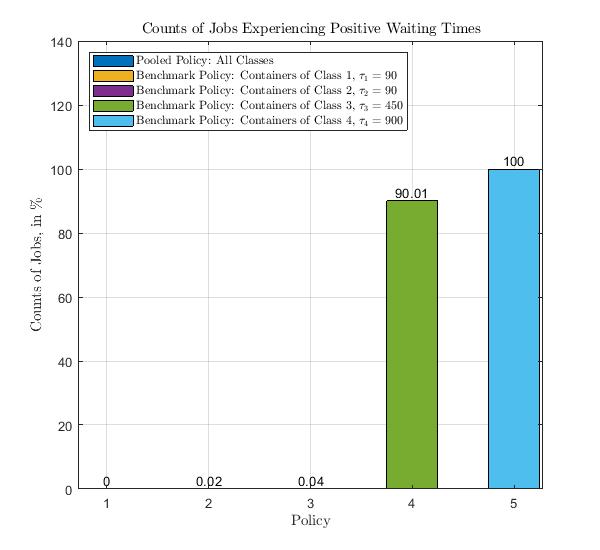}%
       \label{waitCountStat}%
          }
\end{figure}

In addition, we investigate class-specific distributions of service delay to verify if the pre-specified SLAs are respected in simulations. We show in Figure~\ref{survWaitstat} that dedicated to job-classes capacity may result in poor performance (recall that under the pooled policy, all jobs access service immediately). Under the benchmark policy, jobs with lower tolerance for delays (i.e., 90, 90, and 450 minutes)
meet their SLAs. However,
the containers of class 4 remain in the system indefinitely. According to Figure~\ref{waitCountStat}, while service of $0.02\%$ and $0.04\%$ of jobs with stricter SLAs ($\tau_1=90$, $\tau_2=90$, respectively) is delayed, approximately 90\% of containers of class 3 enter a congested system. 
\subsection{Discrete Event Simulation: Containers with Time-varying Demand}\label{sec:simTimeV}
\begin{figure}[h] 
\caption{Time-varying Setting: Dynamics of the Total Capacity Allocation over a Typical Day.}
\label{allocnonstat}
     \subfloat[]{%
        \includegraphics[width=0.53\textwidth]{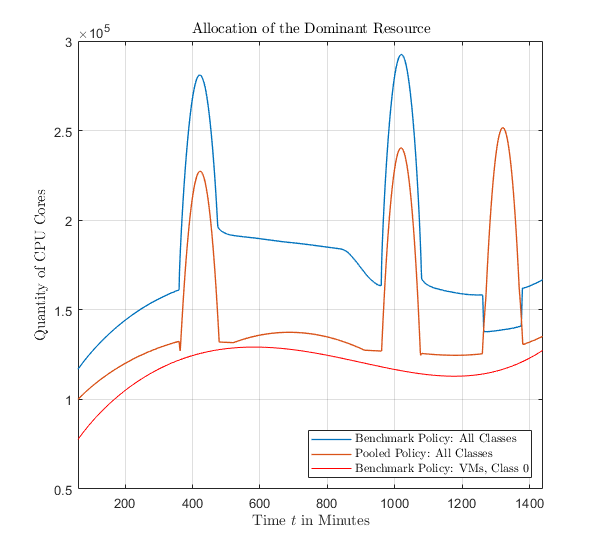}%
        \label{allocCPUnonstat}%
         }%
     \hfill%
     \subfloat[]{%
        \includegraphics[width=0.53\textwidth]{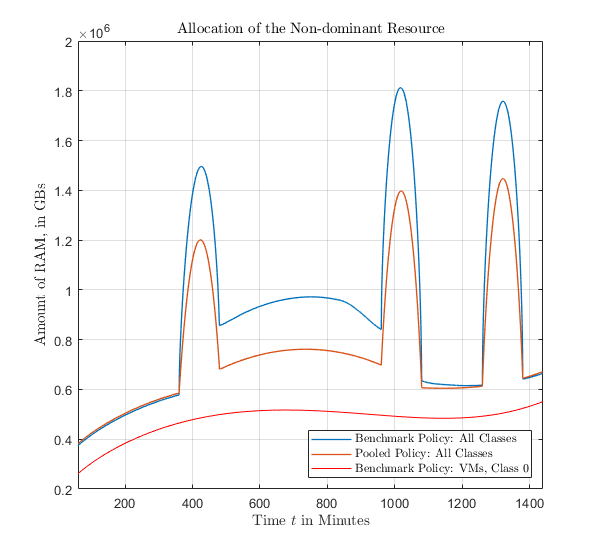}%
        \label{allocRAMnonstat}%
         }
\end{figure}

In this setting, the amount that demand varies with time is driven by the magnitude of spikes observed during the peak hours of containers arriving to the system over the morning, evening and night times (see Figure~\ref{contTV}). By conducting similar tests to Section~\ref{sec:simSta},
we find that, in the time-varying setting, our policy outperforms the benchmark once again. 
In particular, we observe qualitatively similar results to the near-stationary setting in Figure~\ref{allocnonstat}, i.e., our approach provisions significantly less capacity at most times over a typical day. However, influenced by class-4 containers ($\tau_4=900$), the pooled policy assigns more CPU cores than the benchmark at the end of the day. This is because the workload generated by VMs dominates the rest of job classes, and, thus, weight $\omega_0$ associated with the service quality that is required by VMs admits a relatively large value (see Section~\ref{sec:capacity}). In contrast, due to the high tolerance of delays of class-4 containers, the benchmark policy neglects them at the time of arrival and serves them during the following day, instead.

\begin{figure}[h] 
\caption{Time-varying Setting: Distributions of the Capacity Utilization over a Typical Day.}
\label{utilnonstat}
     \subfloat[]{%
      \includegraphics[width=0.53\textwidth]{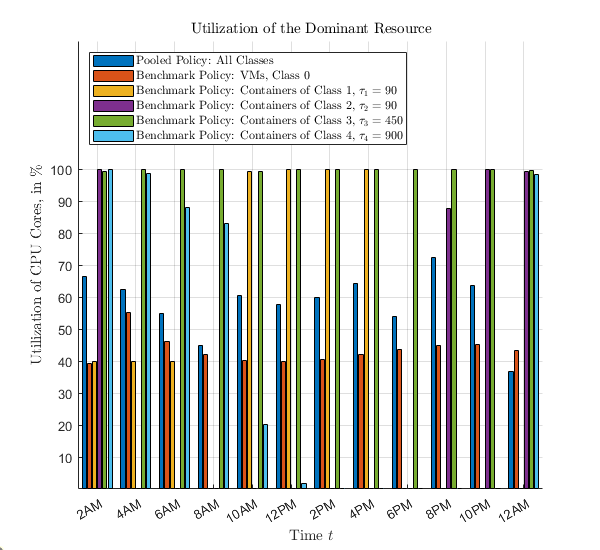}%
        \label{utilCPUallnonStat}%
         }%
     \hfill%
     \subfloat[]{%
        \includegraphics[width=0.53\textwidth]{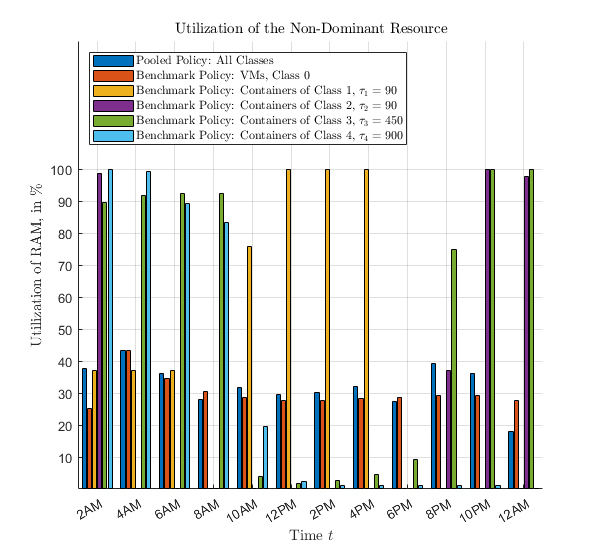}%
        \label{utilRAMallnonStat}%
         }
\end{figure}

Further, according to Figure~\ref{utilnonstat}, the dynamics of service utilization favour the pooled policy as it induces less pronounced resource idling compared to the benchmark (see the capacity reserved for VMs, for instance). Similarly to the near-stationary setting, all jobs access service immediately under the pooled policy.
In contrast, under the benchmark, classes with peak arrivals in the morning, evening, and night experience congestion, and containers with arrivals spread over the daytime observe a busy system at all times. According to the dynamics of RAM utilization in Figure~\ref{utilRAMallnonStat}, a shortage of memory 
may cause delays to containers arriving in the morning. 
 We also observe that there is a time lag between the intensity of arrivals and the system utilization. For instance, morning arrivals increase the cloud workload in the afternoon. 
 
\begin{figure}[h] 
\caption{Time-varying Setting: (a) Likelihood of  Delays; (b) Counts of Delayed Jobs.}
\label{Waitnonstat}
      \subfloat[]{%
        \includegraphics[width=0.53\textwidth]{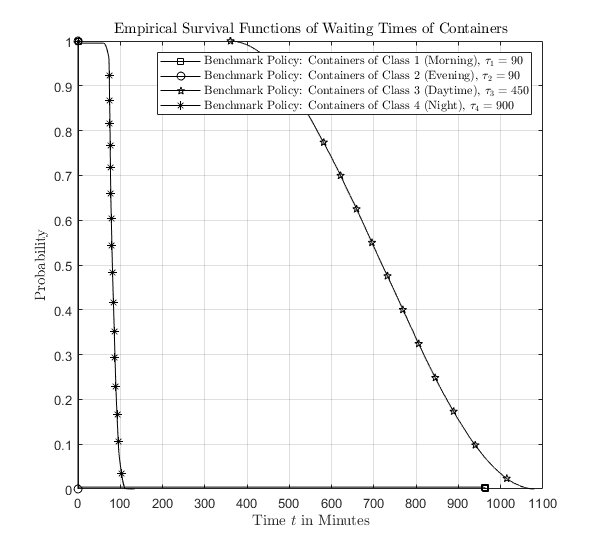}%
        \label{survWaitnonstat}
          }%
      \hfill%
      \subfloat[]{%
       \centering 
            \includegraphics[width=0.53\textwidth]{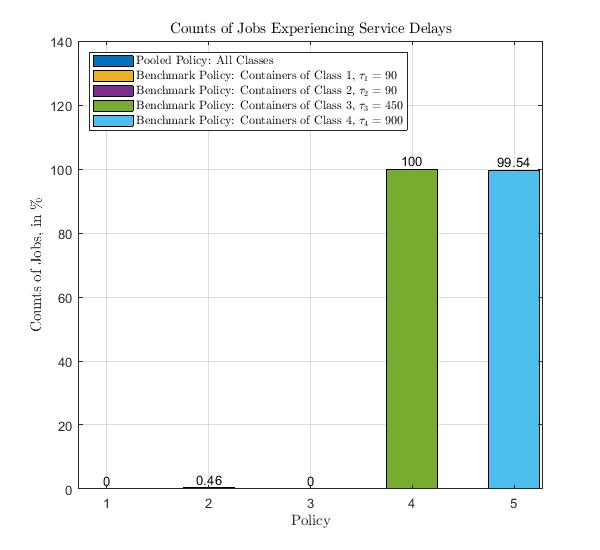}%
            \label{waitCountNonstat}
          }
\end{figure}

Finally, 
in Figure~\ref{survWaitnonstat}, we show that under the benchmark policy,
containers with most variability in demand wait for service significantly less than their pre-specified thresholds (i.e., 90, 90, and 900 minutes), and jobs arriving more uniformly over the daytime do not meet their SLA. 
In addition, similar to Section~\ref{sec:simSta}, few containers of class 1 and 2 experience service delays, while nearly all jobs of class 3 and 4 join a queue in order to get serviced (see Figure~\ref{waitCountNonstat}). 

\section{Discussion and Conclusions}\label{sec:Conc}
In this paper, we analyze a cloud computing service with parallel processing, heterogeneous workload with batches (i.e., VMs and containers), and multiple resources one of which is considered dominant. If VMs do not immediately begin service, they leave the system, while containers are infinitely patient and join the queue if needed. However, the duration of their wait must respect a set of pre-specified SLAs. All jobs in our system require multiple types of resources to get serviced and release them at service completion. Exact analysis of this system with time-varying arrivals is intractable. Thus, we develop a diffusion model 
of the corresponding stochastic offered load (SOL). In our subsequent analysis 
we construct a stochastic approximation of the SOL in the system with parallel processing. Then, we propose a three-stage procedure that employs this approximation and modifies the time-varying square-root-staffing rule to make capacity allocation decisions. 

Our methodology overcomes challenges associated with the analysis of queues with parallel processing, multiple types of resources, and heterogeneous jobs that are typically analysed in stochastic simulations \citep{zychlinski2020scheduling}. Leveraging the approximation of the SOL derived in the $\infty$ stage, we apply an intuitive aggregation rule to individual workloads in order to analyze multi-class service systems with flexible resources (the $\alpha$ and $\epsilon$ stages). Because our approximation of the SOL has a closed-form expression for a large class of arrival patterns, the approach is general and remains tractable for arrival functions that admit basic algebraic operations, i.e., 
demand complexity is not inherited by the capacity allocation stages of our method. In addition, as we link the capacity levels that respect heterogeneous SLAs of jobs at each time instant to the percentile curves of the SOL available in closed-form, our approach is computationally efficient - albeit, under some assumptions - such as jobs belonging to a single batch have identical durations of service.

Our analysis suggests that in settings where demand is near-stationary or highly time-varying, transitioning to a resource sharing from reserving capacity is beneficial. That is, services with pooled capacity may achieve better performance while provisioning fewer resources. Specifically, we show that our policy does not violate the SLAs and requires at least 20\% less capacity than the 
benchmark, i.e., savings are achievable together with a positive service impact.

Further, results from our simulation study suggest that there is a relatively high fidelity between the SOL approximation and the workload induced by the data. 
In addition, the SOL-based weights that we incorporate in our heuristic policy improve its overall performance in the near-stationary setting, which results in meeting the SLAs and maintaining high utilization of the dominant resource (see Figure~\ref{utilCPUallStat}). To the contrary, the benchmark policy that dedicates resources to job classes 
completely neglects jobs with the largest tolerance for service delays.

We also find that our heuristic handles variability in demand better than the benchmark. In particular, in Figure~\ref{allocCPUnonstat} and \ref{utilCPUallnonStat}, the heuristic policy assigns sufficient resources to process containers with peak arrivals during the night time and does not neglect jobs of other classes (see the time interval from 8pm to 12am). However, the capacity levels set by the benchmark are insufficient, which results in a violation of the SLA of containers whose arrivals are spread over the daytime.      
Nevertheless, our pooled policy is influenced by a high proportion of workload contributed to the system by VMs. Their intolerance for delays results in over-provisioning of resources ensuring that service denial does not occur in the stochastic regime. Because capacity unused by VMs can be employed by containers, the utilization of a cloud system operating under such policy remains smaller than one at all times, which guaranties that all arrivals can access service immediately.  

Similarly to \cite{FurmanDiamant2021}, our study suggests that cloud computing providers require large amounts of capacity to satisfy challenging service level agreements, which results in the resource idling \citep[see][for instance]{Forbes,FLEXERA}. To this end, we encourage cloud service operators to pursue configurations that do not dedicate resources but share them amongst jobs in a flexible manner.  Discussions with our partner organization indicate that they share these conclusions, i.e., there is a general consensus amongst major cloud computing providers that resource pooling constitutes a superior design and must be pursued if possible.    

Finally, our approach applies to cases where managers are concerned
with the long-term behavior of the system because cloud computing
services operate 24 hours a day and quickly achieve their steady state. 
Further, as variability in the system workload increases farther away from its initial state, our approach tends to assign less capacity at the beginning of the day than in the late evening. Thus, we recommend that cloud operators interpret the proposed capacity levels over the first hour of a typical 24-hour period with greater care. In addition, our method requires providers to be familiar with dynamics of the offered load in their service system. This information is required to set the weights in the heuristic policy and can be obtained from the historical data. 
\bibliographystyle{informs2014} 
\bibliography{references}

\ECSwitch
\renewcommand{\thetable}{\arabic{table}}  
\ECHead{Appendices and Proofs of Statements}
\section{Simulation Parameters}\label{appedA}
\vspace{5pt}
\noindent \scriptsize Table~\ref{nearStat} and \ref{timeVar} present the arrival functions of 4 classes of containers in a near stationary and time-varying setting, respectively.\\

\noindent
Table~\ref{QueParST} and \ref{QueParTV} provide a summary of the generated data that we employ to parameterize our experiments. Specifically, we generate 10 sets of data using bootstrap distributions of service times, fitted probability mass functions of capacity requirements, and non-homogeneous Poisson batch arrival processes with mean rates presented in Table~\ref{nearStat} and \ref{timeVar}. For example, 666.424 in Column 2, is the grand average of mean service times of VMs observed in 10 stochastic experiments in a near-stationary setting.
\begin{table}[h]
	\begin{minipage}{.5\linewidth}
  	\caption{Near-stationary Containers}\label{nearStat}
      \scalebox{0.9}{
        \begin{tabular}{@{}ll@{}}
        \toprule
        \multicolumn{1}{c}{} & \multicolumn{1}{c}{Batch  Arrival Function}                                               \\ \midrule
        Container 1          & $\lambda_1(t) = -\frac{1}{3,568,896}t^2-\frac{155}{297,408}t+\frac{191,875}{13,941}$    \\[0.1cm]
        Container 2          & $\lambda_2(t) = -\frac{19}{81,544,320}t^2+\frac{19}{28,512}t+\frac{729,125}{56,628}$   \\[0.1cm]
        Container 3          & $\lambda_3(t) = -\frac{19}{92,314,944}t^2+\frac{1235}{5,769,684}t+\frac{25,353,125}{1,923,228}$ \\ [0.1cm]
        Container 4          & $\lambda_4(t) = -\frac{19}{85,578,624}t^2+\frac{29,165}{21,394,656}t+\frac{33,071,875}{2,674,332}$ \\[0.1cm] \bottomrule
        \end{tabular}
        }
	\end{minipage}
\begin{minipage}{.5\linewidth}
  	\caption{Time-varying Containers}\label{timeVar}
      \centering
      \scalebox{0.9}{
	    \begin{tabular}{@{}ll@{}}
        \toprule
        \multicolumn{1}{c}{} & \multicolumn{1}{c}{Batch  Arrival Function}                                          \\ \midrule
        Container 1          & $\lambda_1(t) = -\frac{19}{288}t^2+\frac{665}{12}t-11,400$    \\[0.1cm]
        Container 2          & $\lambda_2(t) = -\frac{19}{288}t^2+\frac{1615}{12}t-68,400$   \\[0.1cm]
        Container 3          & $\lambda_3(t) = -\frac{19}{62,208}t^2+\frac{95}{216}t-118.75$ \\ [0.1cm]
        Container 4          & $\lambda_4(t) = -\frac{19}{288}t^2+\frac{1045}{6}t-114,712.5$ \\[0.16cm] \bottomrule
        \end{tabular}
        }
	\end{minipage}
	\vspace{0.3cm} 
	\caption{Average Queueing Parameters: Near-stationary Setting}\label{QueParST}
\centering
\scalebox{0.8}{
\begin{tabular}{@{}lccccc@{}}
\toprule
                                       \thead{Column \bf{1}\\ { }}       &  \thead{Column \bf{2}\\VMs, $i=0$} & \thead{Column \bf{3}\\Containers, $i=1$} & \thead{Column \bf{4}\\Containers, $i=2$} & \thead{Column \bf{5}\\Containers, $i=3$} & \thead{Column \bf{6}\\Containers, $i=4$} \\ \midrule
Mean Service Time, $1/\mu_i$, in Minutes      & 666.424    & 89.589            & 89.439            & 89.744            & 89.417            \\
St. Dev. Service Time, $\sigma_i$, in Minutes & 1,681      & 172.306           & 172.170           & 172.293           & 171.969           \\ \midrule
Mean Batch Size, $v_i$                        & 1.439      & 1.439             & 1.438             & 1.438             & 1.440             \\
St. Dev. Batch Size, $\beta_i$                & 1.344      & 1.338             & 1.341             & 1.340             & 1.340             \\ \midrule
Mean CPU-core Req., $r_{i1}$                  & 4.108      & 4.103             & 4.110             & 4.010             & 4.099             \\
St. Dev. CPU-core Req., $\delta_{i1}$         & 7.812      & 7.785             & 7.802             & 7.785             & 7.763             \\ \midrule
Mean RAM Req., $r_{i2}$, in GBs               & 11.542     & 11.488            & 11.504            & 11.495            & 11.510            \\
St. Dev. RAM Req., $\delta_{i2}$, in GBs      & 34.505     & 33.946            & 34.255            & 34.528            & 34.288            \\ \bottomrule
\end{tabular}
}
\vspace{0.5cm} 
\caption{Average Queueing Parameters: Time-varying Setting}\label{QueParTV}
\centering
\scalebox{0.8}{
\begin{tabular}{@{}lccccc@{}}
\toprule
                                       \thead{Column \bf{1}\\ { }}       & \thead{Column \bf{2}\\VMs, $i=0$} & \thead{Column \bf{3}\\Containers, $i=1$}  & \thead{Column \bf{4}\\Containers, $i=2$} & \thead{Column \bf{5}\\Containers, $i=3$} & \thead{Column \bf{5}\\Containers, $i=4$} \\ \midrule
Mean Service Time, $1/\mu_i$, in Minutes      & 667.588    & 89.671            & 89.412            & 89.631            & 90.043            \\
St. Dev. Service Time, $\sigma_i$, in Minutes & 1,682      & 172.326           & 172.547           & 172.321           & 171.691           \\ \midrule
Mean Batch Size, $v_i$                        & 1.439      & 1.440             & 1.439             & 1.436             & 1.438             \\
St. Dev. Batch Size, $\beta_i$                & 1.342      & 1.354             & 1.339             & 1.334             & 1.335             \\ \midrule
Mean CPU-core Req., $r_{i1}$                  & 4.120      & 4.106             & 4.093             & 4.106             & 4.104             \\
St. Dev. CPU-core Req., $\delta_{i1}$         & 7.831      & 7.805             & 7.746             & 7.819             & 7.782             \\ \midrule
Mean RAM Req., $r_{i2}$, in GBs               & 11.497     & 11.469            & 11.549            & 11.508            & 11.520            \\
St. Dev. RAM Req., $\delta_{i2}$, in GBs      & 34.278     & 33.587            & 34.747            & 34.130            & 34.884            \\ \bottomrule
\end{tabular}
}
\end{table}
\newpage
\section*{Summary of the Results: System Utilization}\label{appedB}
\vspace{5pt}
Tables~\ref{CPUutilSta} - \ref{RAMutilSta} and Tables~\ref{CPUutilnonSta} - \ref{RAMutilnonSta} accompany the bar plots of the system utilization in Section~\ref{sec:NumStu} in the near-stationary and time-varying setting, respectively. In these tables, the first column specifies a capacity allocation policy. For example, ``Pooled policy'' refers to the proposed heuristic, while ``VMs'' represents resources dedicated to virtual machines under the benchmark policy.
The remaining columns correspond to the values over a weekday with 2-hour intervals. We present the largest values of the utilization below 90\% in bold, while the instances when the system operates closer to full capacity are highlighted in red.  
\begin{table}[h]
\caption{System Utilization of the Dominant Resource (\%): Near-stationary Setting}\label{CPUutilSta}
\centering
\scalebox{0.7}{
\begin{tabular}{@{}lcccccccccccc@{}}
\toprule
               \thead{Column \bf{1}\\ { }}  & \thead{Column \bf{2}\\ {2 am}}            & \thead{Column \bf{3}\\ {4 am}}           & \thead{Column \bf{4}\\ {6 am}}            & \thead{Column \bf{5}\\ {8 am}}            & \thead{Column \bf{6}\\ {10 am}}           & \thead{Column \bf{7}\\ {12 am}}            & \thead{Column \bf{8}\\ {2 pm}}             & \thead{Column \bf{9}\\ {4 pm}}            & \thead{Column \bf{10}\\ {6 pm}}            & \thead{Column \bf{11}\\ {8 pm}}            & \thead{Column \bf{12}\\ {10 pm}}          & \thead{Column \bf{13}\\ {12 pm}}           \\ \midrule
Pooled Policy & \textbf{60.256} & \textbf{83.712} & \textbf{71.311} & \textbf{65.396} & \textbf{62.666}  & \textbf{61.959}  & \textbf{62.432}  & \textbf{63.683}  & \textbf{64.921} & \textbf{65.472} & \textbf{65.294} & \textbf{63.384} \\
VMs              & 40.369          & 57.154          & 47.340          & 42.963          & 41.279           & 40.949           & 42.037           & 43.436           & 45.054          & 46.136          & 46.224          & 44.198          \\
Container 1      & 37.311          & 79.653          & 66.305          & 58.741          & 54.745           & 50.139           & 46.860           & 44.826           & 42.819          & 41.204          & 39.596          & 38.868          \\
Container 2      & 39.150          & 76.624          & 63.365          & 55.764          & 49.881           & 47.362           & 44.324           & 42.612           & 41.263          & 40.595          & 38.818          & 38.277          \\
Container 3      & \color{red}\textbf{99.970}          & \color{red}\textbf{99.736}          & \color{red}\textbf{99.779}          & \color{red}\textbf{99.566}          & {\color{red}\textbf{100.000}} & {\color{red}\textbf{100.000}} & {\color{red}\textbf{100.000}} & {\color{red}\textbf{100.000}} & \color{red}\textbf{99.979}          & \color{red}\textbf{99.950}          & \color{red}\textbf{99.981}          & \color{red}\textbf{99.975}          \\
Container 4      & 0.000           & 0.000           & 0.000           & 0.000           & 0.000            & 0.000            & 0.000            & 0.000            & 0.000           & 0.000           & 0.000           & 0.000           \\ \bottomrule
\end{tabular}
}
\vspace{0.5cm} 
\caption{System Utilization of the Non-dominant Resource: Near-stationary Setting}\label{RAMutilSta}
\centering
\scalebox{0.7}{
\begin{tabular}{@{}lcccccccccccc@{}}
\toprule
               \thead{Column \bf{1}\\ { }}  & \thead{Column \bf{2}\\ {2 am}}            & \thead{Column \bf{3}\\ {4 am}}           & \thead{Column \bf{4}\\ {6 am}}            & \thead{Column \bf{5}\\ {8 am}}            & \thead{Column \bf{6}\\ {10 am}}           & \thead{Column \bf{7}\\ {12 am}}            & \thead{Column \bf{8}\\ {2 pm}}             & \thead{Column \bf{9}\\ {4 pm}}            & \thead{Column \bf{10}\\ {6 pm}}            & \thead{Column \bf{11}\\ {8 pm}}            & \thead{Column \bf{12}\\ {10 pm}}          & \thead{Column \bf{13}\\ {12 pm}}            \\ \midrule
Pooled Policy & \textbf{28.600} & \textbf{48.715} & \textbf{39.868} & \textbf{35.614} & \textbf{33.272} & \textbf{32.200} & \textbf{31.711} & \textbf{31.696} & \textbf{31.598} & \textbf{31.810} & \textbf{31.201} & \textbf{30.397} \\
VMs              & 24.953          & 42.453          & 33.905          & 30.160          & 28.266          & 27.592          & 27.806          & 28.263          & 28.949          & 29.393          & 29.215          & 27.756          \\
Container 1      & 9.679           & 25.779          & 19.689          & 17.189          & 15.262          & 13.629          & 12.113          & 11.259          & 10.451          & 10.598          & 9.804           & 9.966           \\
Container 2      & 10.271          & 26.412          & 19.626          & 15.829          & 14.027          & 12.648          & 11.430          & 10.978          & 10.607          & 10.599          & 10.017          & 10.038          \\
Container 3      & 10.998          & 18.393          & 9.040           & 3.993           & 7.701           & 10.320          & 13.249          & 15.531          & 17.331          & 15.661          & 14.134          & 12.923          \\
Container 4      & 0.000           & 0.000           & 0.000           & 0.000           & 0.000           & 0.000           & 0.000           & 0.000           & 0.000           & 0.000           & 0.000           & 0.000           \\ \bottomrule
\end{tabular}
}
\vspace{0.5cm} 
\caption{System Utilization of the Dominant Resource (\%): Time-varying Setting}\label{CPUutilnonSta}
\centering
\scalebox{0.7}{
\begin{tabular}{@{}lcccccccccccc@{}}
\toprule
               \thead{Column \bf{1}\\ { }}  & \thead{Column \bf{2}\\ {2 am}}            & \thead{Column \bf{3}\\ {4 am}}           & \thead{Column \bf{4}\\ {6 am}}            & \thead{Column \bf{5}\\ {8 am}}            & \thead{Column \bf{6}\\ {10 am}}           & \thead{Column \bf{7}\\ {12 am}}            & \thead{Column \bf{8}\\ {2 pm}}             & \thead{Column \bf{9}\\ {4 pm}}            & \thead{Column \bf{10}\\ {6 pm}}            & \thead{Column \bf{11}\\ {8 pm}}            & \thead{Column \bf{12}\\ {10 pm}}          & \thead{Column \bf{13}\\ {12 pm}}           \\ \midrule
Pooled Policy & \textbf{66.531}  & \textbf{62.351}  & \textbf{54.965}  & \textbf{44.964}  & \textbf{60.426} & \textbf{57.655} & \textbf{59.824}  & \textbf{64.315}  & \textbf{54.055}  & \textbf{72.419} & \textbf{63.468} & \textbf{36.849} \\
VMs              & 39.347           & 55.095           & 45.989           & 41.955           & 40.046          & 39.851          & 40.570           & 42.171           & 43.706           & 44.995          & 45.033          & 43.219          \\
Container 1      & 39.844           & 39.844           & 39.844           & 0.000            & \color{red}\textbf{99.223}          & \color{red}\textbf{99.972}          & \color{red}\textbf{99.956}           & \color{red}\textbf{99.993}           & 0.000            & 0.000           & 0.000           & 0.000           \\
Container 2      & \color{red}\textbf{99.913}           & 0.091            & 0.091            & 0.091            & 0.091           & 0.091           & 0.091            & 0.091            & 0.000            & 87.603          & \color{red}\textbf{99.949}          & \color{red}\textbf{99.244}          \\
Container 3      & \color{red}\textbf{99.110}           & {\color{red}\textbf{100.000}} & {\color{red}\textbf{100.000}} & {\color{red}\textbf{100.000}} & \color{red}\textbf{99.300}          & \color{red}\textbf{99.902}          & {\color{red}\textbf{100.000}} & {\color{red}\textbf{100.000}} & {\color{red}\textbf{100.000}} & \color{red}\textbf{99.836}          & \color{red}\textbf{99.881}          & \color{red}\textbf{99.555}          \\
Container 4      & {\color{red}\textbf{100.000}} & \color{red}\textbf{98.656}           & 87.908           & 82.828           & 20.120          & 1.874           & 0.287            & 0.287            & 0.287            & 0.287           & 0.287           & \color{red}\textbf{98.286}          \\ \bottomrule
\end{tabular}
}
\vspace{0.5cm} 
\caption{System Utilization of the Non-dominant Resource (\%): Time-varying Setting}\label{RAMutilnonSta}
\centering
\scalebox{0.7}{
\begin{tabular}{@{}lcccccccccccc@{}}
\toprule
               \thead{Column \bf{1}\\ { }}  & \thead{Column \bf{2}\\ {2 am}}            & \thead{Column \bf{3}\\ {4 am}}           & \thead{Column \bf{4}\\ {6 am}}            & \thead{Column \bf{5}\\ {8 am}}            & \thead{Column \bf{6}\\ {10 am}}           & \thead{Column \bf{7}\\ {12 am}}            & \thead{Column \bf{8}\\ {2 pm}}             & \thead{Column \bf{9}\\ {4 pm}}            & \thead{Column \bf{10}\\ {6 pm}}            & \thead{Column \bf{11}\\ {8 pm}}            & \thead{Column \bf{12}\\ {10 pm}}          & \thead{Column \bf{13}\\ {12 pm}}            \\ \midrule
Pooled Policy & \textbf{37.521} & \textbf{43.287} & 35.938          & 28.052          & \textbf{31.652} & \textbf{29.490} & \textbf{30.071} & \textbf{31.879} & 27.399          & 39.301          & \textbf{36.194} & 17.874          \\
VMs              & 25.119          & 43.274          & 34.379          & \textbf{30.416} & 28.606          & 27.693          & 27.719          & 28.267          & \textbf{28.692} & 29.278          & 29.112          & \textbf{27.641} \\
Container 1      & 36.969          & 36.969          & \textbf{36.969} & 0.000           & 75.767          & \color{red}\textbf{99.973}          & \color{red}\textbf{99.946}          & \color{red}\textbf{99.991}          & 0.000           & 0.000           & 0.000           & 0.000           \\
Container 2      & \color{red}\textbf{98.720}          & 0.044           & 0.044           & 0.044           & 0.044           & 0.044           & 0.044           & 0.044           & 0.000           & 36.911          & \color{red}\textbf{99.965}          & \color{red}\textbf{97.592}          \\
Container 3      & 89.549          & \color{red}\textbf{91.575}          & \color{red}\textbf{92.223}          & \color{red}\textbf{92.283}          & 3.795           & 1.734           & 2.806           & 4.522           & 9.128           & \textbf{74.826} & \color{red}\textbf{99.819}          & \color{red}\textbf{99.931}          \\
Container 4      & \color{red}\textbf{99.989}          & \color{red}\textbf{99.154}          & 89.110          & 83.289          & 19.547          & 2.203           & 1.132           & 1.132           & 1.132           & 1.132           & 1.132           & 0.018           \\ \bottomrule
\end{tabular}
}
\end{table}
\newpage
\normalsize
\section*{Proofs of the results}\label{appedC}
\subsection*{Proof of Lemma \ref{lemma:offeredLoad}}
\proof{Proof}
Let $M_{in}$ be a random variable representing type-$n$ capacity requirement of a class-$i$ job. Consider an infinite capacity system $\mathcal{QN}^e_\infty$ and write out its type-$n$ utilization by class-$i$ jobs
\[
X^{e\infty}_{in}(t) = \sum_{k = 1}^{N_{i}(t)} M_{ink},
\]
where, for fixed $i$ and $n$, $M_{ink}$ and $M_{in}$ are i.i.d. for any $k$, and $N_{i}(t)$ is a type-$n$ offered load generated by class-$i$ jobs in $\mathcal{QN}^{\mathbbm{1}e}_\infty$.
Recall that each job in $\mathcal{QN}^e_\infty$ is a batch of jobs  observed in $\mathcal{QN}_\infty$, thus, for a random variable $V_i$ denoting a batch size of class-$i$ jobs, we have that
\[
X^{\infty}_{in}(t) = \sum_{k = 1}^{N_{i}(t)} \sum_{l = 1}^{V_{ik}}M_{inkl},
\]
where, for fixed $i$ and $n$, $M_{inkl}$ and $M_{in}$ are i.i.d. and $V_{ik}$ and $V_{p}$ are i.i.d. for any $k$ and $l$.
Let $B_{ink}\coloneqq \sum_{l = 1}^{V_{ik}}M_{inkl}$, then because $\{V_{ik}\}_{k}$ are i.i.d., and $\{M_{inkl}\}_{kl}$ are i.i.d. for any $k$ and $l$, $\{B_{ink}\}_k$ are also i.i.d. Further, because $\mathbb{E}[V_i]<\infty$, $Var[V_i]<\infty$, $\mathbb{E}[M_{in}]<\infty$, and $Var[M_{in}]<\infty$, both expectation and variance of $B_{ink}$ are finite for any $k$. Thus, for fixed $i$ and $n$, because $N_{i}(t)$ is a non-homogeneous Poisson process, and $B_{ink}$ are i.i.d. for any $k$,  we have that
\[
X^{\infty}_{in}(t) = \sum_{k = 1}^{N_{i}(t)} B_{ink}
\]
is a compound non-homogeneous Poisson process, and its expectation and variance are well known:
\[
\mathbb{E}[X^{\infty}_{in}(t)] = \mathbb{E}[N_i(t)] \mathbb{E}[B_{in1}] = m_i(t)\mathbb{E}[B_{in1}]
\]
and
\[
Var[X^{\infty}_{in}(t)] = m_i(t)\left(Var[B_{in1}]+\mathbb{E}[B_{in1}]^2\right). 
\]
To complete the proof, we derive $\mathbb{E}[B_{in1}]$ and $Var[B_{in1}]$.  $\mathbb{E}[B_{in1}]$ immediately follows from Wald's equation
\[
\mathbb{E}[B_{in1}] = \mathbb{E}[M_{in11}+M_{in12}+...+M_{in1V_{ik}}] = \mathbb{E}[V_{ik}]\mathbb{E}[M_{in}] = v_ir_{in}.
\]
Then, we apply the law of total variance to obtain $Var[B_{in1}]$
\begin{align*}
Var[B_{in1}] &= \mathbb{E}[Var[B_{in1}|V_{i1}]] + Var[\mathbb{E}[B_{in1}|V_{i1}]]\\
&=\mathbb{E}[V_{i1}Var[M_{in}]] + Var[V_{i1}\mathbb{E}[M_{in}]]\\
&=Var[M_{in}]\mathbb{E}[V_{i1}] + \mathbb{E}[M_{in}]^2Var[V_{i1}]\\
&=\delta_{in}^2v_i + r_{in}^2\beta_i^2. 
\end{align*}
Therefore,
\[
\mathbb{E}[X^{\infty}_{in}(t)] = v_ir_{in}m_i(t),
\quad
Var[X^{\infty}_{in}(t)] = m_i(t)\left(\delta_{in}^2v_i + r_{in}^2\beta_i^2+v_i^2r_{in}^2\right). 
\]
\hfill\qed
\endproof
\subsection*{Proof of Proposition \ref{Pro:diffusion}}
\proof{Proof}
We scale the arrival process to $\mathcal{QN}^e_{\infty}$ up by factor $\xi$, i.e., the scaled intensity of class-$i$ batch arrivals equals  $\xi\lambda_{i}(t)$, and the scaled $X^{\infty}_{in}(t)$ is denoted by $X^{\xi\infty}_{in}(t)$.
Applying the functional law of large numbers, we have the following fluid limit
\[
lim_{\xi\to\infty} \frac{X^{\xi\infty}_{in}(t)}{\xi} = \mathbb{E}[X^{\infty}_{in}(t)].
\]
Let $\bar{X}^{\infty}_{in}(t;\xi)=\frac{1}{\xi}(X^{\xi\infty}_{in}(t))$. Then, by the functional central limit theorem, 
\[
\lim_{\xi\to{\infty}}\sqrt{\xi}(\bar{X}^{\infty}_{in}(t;\xi) - \mathbb{E}[X^{\infty}_{in}(t)])
= \sqrt{Var[X^{\infty}_{in}(t)]}W_{in}(t),
\]
where $W_{in}(t)$ is the standard Wiener process. 

We have that $dX^{\infty}_{in}(t) \simeq \left(\mathbb{E}[X^{\infty}_{in}(t)]\right)^{\prime}dt + \sqrt{Var[X^{\infty}_{in}(t)]}dW_{in}(t)$.
By integrating both sides, for $t>0$, the desired result is obtained,
\begin{align*}X^{\infty}_{in}(t)&\simeq \hat{X}^{\infty}_{in}(t)= X^{\infty}_{in}(0) - \mathbb{E}[X^{\infty}_{in}(0)]+ \mathbb{E}[X^{\infty}_{in}(t)]
+ \sqrt{Var[X^{\infty}_{in}(t)]}W_{in}(t)\\  &= X^{\infty}_{in}(0) - v_ir_{in}m_i(0) + v_ir_{in}m_i(t) + \sqrt{m_i(t)\left(\delta_{in}^2v_i + r_{in}^2\beta_i^2+v_i^2r_{in}^2\right)}W_{in}(t).
\end{align*}
\hfill\qed
\endproof
\subsection*{Proof of Corollary \ref{co:bounds}}
\proof{Proof}
Because $W_{np}(t)\sim N(0,t)$, we re-write \eqref{wiener:approx}
\[
\hat{X}^{\infty}_{in}(t)= X^{\infty}_{in}(0) - \mathbb{E}[X^{\infty}_{in}(0)]+ \mathbb{E}[X^{\infty}_{in}(t)]
+ \sqrt{tVar[X^{\infty}_{in}(t)]}Z,
\]
where $Z$ is a standard normal random variable.

Clearly, $\hat{X}^{\infty}_{in}(t)\sim N(X^{\infty}_{in}(0) - \mathbb{E}[X^{\infty}_{in}(0)]+ \mathbb{E}[X^{\infty}_{in}(t)],tVar[X^{\infty}_{in}(t)])$. The result in \eqref{prcAlpha} immediately follows.\hfill\qed
\endproof
\subsection*{Proof of Corollary \ref{co:bounds_pooled}}
\proof{Proof}
Because $\hat{X}^{\infty}_{in}(t)$ are independent stochastic processes for $i\in\mathcal{I}$ and $n\in\mathcal{N}$, the result follows immediately from Corollary~\ref{co:bounds}.\hfill\qed
\endproof
\subsection*{Proof of Corollary \ref{co:siloPol}}
\proof{Proof}
Follows from Corollary~\ref{co:bounds} and Proposition~\ref{pro:Capacity}.\hfill\qed
\endproof
\subsection*{Proof of Proposition \ref{pro:Capacity}}
\proof{Proof}
\begin{enumerate}
    \item For a fixed $t$, we consider \ref{def_alphaSLA}
    \noindent$\mathbb{P}\left(\sum_{i=0}^I    X_{i\tilde{n}}^{\infty}(t)> \tau_i\sum_{j=0}^Iv_jr_{j\tilde{n}}\lambda_j(t)+\nu_{i\tilde{n}}(t)\right) \simeq \mathbb{P}\left(\sum_{i=0}^I    \hat{X}_{i\tilde{n}}^{\infty}(t)> \tau_i\sum_{j=0}^Iv_jr_{j\tilde{n}}\lambda_j(t)+\nu_{i\tilde{n}}(t)\right)$.
Per Corollary~\ref{co:bounds_pooled}, $\sum_{i=0}^I\hat{X}^{\infty}_{i\tilde{n}}(t)\sim N\left(\sum_{i=0}^IX^{\infty}_{i\tilde{n}}(0) - \mathbb{E}[X^{\infty}_{i\tilde{n}}(0)]+ \mathbb{E}[X^{\infty}_{i\tilde{n}}(t)],\sum_{i=0}^ItVar[X^{\infty}_{i\tilde{n}}(t)]\right)$. We have that
\[
\mathbb{P}\left(\sum_{i=0}^I    \hat{X}_{i\tilde{n}}^{\infty}(t)> \tau_i\sum_{j=0}^Iv_jr_{j\tilde{n}}\lambda_j(t)+\nu_{i\tilde{n}}(t)\right) = 
\mathbb{P}\left(Z> \frac{\tau_i\sum_{j=0}^Iv_jr_{j\tilde{n}}\lambda_j(t)+\nu_{i\tilde{n}}(t)-\mathbb{E}\left[\sum_{i=0}^I\hat{X}^{\infty}_{i\tilde{n}}(t)\right]}{\sqrt{\sum_{i=0}^ItVar[X^{\infty}_{i\tilde{n}}(t)]}}\right),
\]
where $Z\sim N(0,1)$. Denote 
\[
\zeta_{i\tilde{n}}(t) = \frac{\tau_i\sum_{j=0}^Iv_jr_{j\tilde{n}}\lambda_j(t)+\nu_{i\tilde{n}}(t)-\sum_{i=0}^I\left(X^{\infty}_{i\tilde{n}}(0) - \mathbb{E}[X^{\infty}_{i\tilde{n}}(0)]+ \mathbb{E}[X^{\infty}_{i\tilde{n}}(t)]\right)}{\sqrt{\sum_{i=0}^Itm_i(t)\left(\delta_{i\tilde{n}}^2v_i + r_{i\tilde{n}}^2\beta_i^2+v_i^2r_{i\tilde{n}}^2\right)}}, 
\]
then, we have $\mathbb{P}\left(\sum_{i=0}^I    \hat{X}_{i\tilde{n}}^{\infty}(t)> \tau_i\sum_{j=0}^Iv_jr_{j\tilde{n}}\lambda_j(t)+\nu_{i\tilde{n}}(t)\right)=1-\Phi(\zeta_{i\tilde{n}}(t))$. We set $1-\Phi(\zeta_{i\tilde{n}}(t)) = \alpha_i$ and solve for $\nu_{i\tilde{n}}(t)$. 
Thus, 
\begin{align}
\hat{\nu}_{i\tilde{n}}(t) &= \Phi^{-1}(1-\alpha_i)\sqrt{\sum_{i=0}^Itm_i(t)\left(\delta_{i\tilde{n}}^2v_i + r_{i\tilde{n}}^2\beta_i^2+v_i^2r_{i\tilde{n}}^2\right)}+ \sum_{i=0}^I\left(X^{\infty}_{i\tilde{n}}(0) - v_ir_{i\tilde{n}}m_i(0) + v_ir_{i\tilde{n}}m_i(t)\right)\notag\\
&-\tau_i\sum_{j=0}^Iv_jr_{j\tilde{n}}\lambda_j(t).
\end{align}
Define 
\[
x_{\cdot{}\tilde{n}}^{\gamma_i}(t) \coloneqq
\Phi^{-1}(\gamma_i)\sqrt{\sum_{i=0}^Itm_i(t)\left(\delta_{i\tilde{n}}^2v_i + r_{i\tilde{n}}^2\beta_i^2+v_i^2r_{i\tilde{n}}^2\right)}+ \sum_{i=0}^I\left(X^{\infty}_{i\tilde{n}}(0) - v_ir_{i\tilde{n}}m_i(0) + v_ir_{i\tilde{n}}m_i(t)\right),
\]
where $1-\alpha_i = \gamma_i$, which completes the proof. 
\hfill\qed
\end{enumerate}
\endproof
\subsection*{Proof of Proposition~\ref{pro:secRes}}
\proof{Proof}
Because $\sum_{i=0}^IX_{in}^{\infty}(t)$ and $\sum_{i=0}^IX_{i\tilde{n}}^{\infty}(t)$ are independent for any $n\ne\tilde{n}$ and time $t\in[0,T]$, using the SOL approximation of $X_{in}^{\infty}(t)$, we require that the following inequality holds 
\[
\mathbb{P}\left(\sum_{i=0}^IX_{in}^{\infty}(t)>s_{in}(t)\right)\approx\mathbb{P}\left(\sum_{i=0}^I\hat{X}_{in}^{\infty}(t)>s_{in}(t)\right)\le \epsilon, \quad \forall i\in\mathcal{I},\quad \forall n\ne \tilde{n}\in\mathcal{N}.
\]
Similarly to Proposition~\ref{pro:Capacity}, we have that 
\[
\hat{s}_{n}(t) = \Phi^{-1}(1-\epsilon)\sqrt{\sum_{i=0}^Itm_i(t)\left(\delta_{i\tilde{n}}^2v_i + r_{i\tilde{n}}^2\beta_i^2+v_i^2r_{i\tilde{n}}^2\right)}+ \sum_{i=0}^I\left(X^{\infty}_{i\tilde{n}}(0) - v_ir_{i\tilde{n}}m_i(0) + v_ir_{i\tilde{n}}m_i(t)\right)=x_{\cdot n}^{\alpha}(t).
\]
Notice that when there is sufficient quantity of the dominant resource, $\hat{s}_{n}(t)$ ensures that non-dominant resources do not increase service delay. However, at times, when  the quantity of the dominant resource is not sufficient, availability of the other resources does not impact service quality. Thus, because $\hat{s}_{\tilde{n}}(t)$ is the solution to \ref{alphaOPT}, we have that $\boldsymbol{\hat{s}}(t)=(\hat{s}_{1}(t), \hat{s}_{2}(t),...,\hat{s}_{n}(t))$
solves \ref{epsilonOPT} and satisfies \ref{def_SLA} as long as $\hat{s}_{\tilde{n}}(t)$ does so, which completes the proof.
\hfill\qed
\endproof
\end{document}